\documentstyle[epsfig,longtable,times]{mn}

\newif\ifAMStwofonts

\def\mabs{$M_{\rm B}$}
\def\muv{$m_{\rm UV}$}
\def\mb{$m_{\rm B}$}

\def\hii{H{\sc ii}}
\def\cbeta{$c_{\rm H\beta}$}
\def\av{$A_{\rm v}$}

\def\tnii{$t_{\rm [N\,{\sc ii}]}$}
\def\doh{$12 + \log(\rm O/H)$}
\def\lno{$\log(\rm N/O)$}
\def\micron{$\mu$m}

\def\kmsmpc{km s$^{-1}$ Mpc$^{-1}$}
\def\cmc{cm$^{-3}$}

\def\ergs{ergs s$^{-1}$}
\def\ergscm{ergs s$^{-1}$ cm$^{-2}$}

\def\msun{M$_{\odot}$}
\def\zsun{Z$_{\odot}$}
\def\halpha{\ifmmode {\rm H{\alpha}} \else $\rm H{\alpha}$\fi}
\def\hbeta{\ifmmode {\rm H{\beta}} \else $\rm H{\beta}$\fi}
\def\oi{[O\,{\sc i}] $\lambda$6300}
\def\oii{[O\,{\sc ii}] $\lambda$3727}
\def\oiii{[O\,{\sc iii}] $\lambda\lambda$4959,5007}
\def\oiiia{[O\,{\sc iii}] $\lambda$4959}
\def\oiiib{[O\,{\sc iii}] $\lambda$5007}
\def\oiiic{[O\,{\sc iii}] $\lambda$4363}
\def\ntwo{[N\,{\sc ii}]}
\def\nii{[N\,{\sc ii}] $\lambda$6584}
\def\Nii{[N\,{\sc ii}] $\lambda\lambda$6548,6584}

\def\sii{[S\,{\sc ii}] $\lambda\lambda$6717+6731}
\def\heii{He\,{\sc ii} $\lambda$4686}
\def\niii{N\,{\sc iii} $\lambda$4640}
\def\ciii{C\,{\sc iii} $\lambda$5696}
\def\civ{C\,{\sc iv} $\lambda$5808-12}
\def\rr23{$R_{\rm 23}$}
\def\oo32{$O_{\rm 32}$}

\ifoldfss
  \newcommand{\rmn}[1] {{\rm #1}}

  \ifCUPmtlplainloaded \else
    \NewTextAlphabet{textbfit} {cmbxti10} {}
    \NewTextAlphabet{textbfss} {cmssbx10} {}
    \NewMathAlphabet{mathbfit} {cmbxti10} {} 
    \NewMathAlphabet{mathbfss} {cmssbx10} {} 
  \fi
  \ifAMStwofonts
    \ifCUPmtlplainloaded \else
      \NewSymbolFont{upmath} {eurm10}
      \NewSymbolFont{AMSa} {msam10}
      \NewMathSymbol{\upi}     {0}{upmath}{19}
      \NewMathSymbol{\umu}     {0}{upmath}{16}
      \NewMathSymbol{\upartial}{0}{upmath}{40}
      \NewMathSymbol{\leqslant}{3}{AMSa}{36}
      \NewMathSymbol{\geqslant}{3}{AMSa}{3E}

      \let\leq=\leqslant \let\leq=\leqslant
      \let\geq=\geqslant \let\geq=\geqslant
    \fi
  \fi
\fi 

\ifnfssone
  \newmathalphabet{\mathit}
  \addtoversion{normal}{\mathit}{cmr}{m}{it}
  \addtoversion{bold}{\mathit}{cmr}{bx}{it}
  \newcommand{\rmn}[1] {\mathrm{#1}}

  \newmathalphabet{\mathbfit} 
  \addtoversion{normal}{\mathbfit}{cmr}{bx}{it}
  \addtoversion{bold}{\mathbfit}{cmr}{bx}{it}
  \newmathalphabet{\mathbfss} 
  \addtoversion{normal}{\mathbfss}{cmss}{bx}{n}
  \addtoversion{bold}{\mathbfss}{cmss}{bx}{n}
  \ifAMStwofonts
    \ifCUPmtlplainloaded \else
      \UseAMStwoboldmath
      \makeatletter
      \new@mathgroup\upmath@group
      \define@mathgroup\mv@normal\upmath@group{eur}{m}{n}
      \define@mathgroup\mv@bold\upmath@group{eur}{b}{n}
      \edef\UPM{\hexnumber\upmath@group}
      \new@mathgroup\amsa@group
      \define@mathgroup\mv@normal\amsa@group{msa}{m}{n}
      \define@mathgroup\mv@bold\amsa@group{msa}{m}{n}
      \edef\AMSa{\hexnumber\amsa@group}
      \makeatother
      \mathchardef\upi="0\UPM19
      \mathchardef\umu="0\UPM16
      \mathchardef\upartial="0\UPM40
      \mathchardef\leqslant="3\AMSa36
      \mathchardef\geqslant="3\AMSa3E

      \let\leq=\leqslant \let\leq=\leqslant
      \let\geq=\geqslant \let\geq=\geqslant
    \fi
  \fi
\fi 

\ifnfsstwo
  \newcommand{\rmn}[1] {\mathrm{#1}}

  \DeclareMathAlphabet{\mathbfit}{OT1}{cmr}{bx}{it}
  \SetMathAlphabet\mathbfit{bold}{OT1}{cmr}{bx}{it}
  \DeclareMathAlphabet{\mathbfss}{OT1}{cmss}{bx}{n}
  \SetMathAlphabet\mathbfss{bold}{OT1}{cmss}{bx}{n}
  \ifAMStwofonts
    \ifCUPmtlplainloaded \else
      \DeclareSymbolFont{UPM}{U}{eur}{m}{n}
      \SetSymbolFont{UPM}{bold}{U}{eur}{b}{n}
      \DeclareSymbolFont{AMSa}{U}{msa}{m}{n}
      \DeclareMathSymbol{\upi}{0}{UPM}{"19}
      \DeclareMathSymbol{\umu}{0}{UPM}{"16}
      \DeclareMathSymbol{\upartial}{0}{UPM}{"40}
      \DeclareMathSymbol{\leqslant}{3}{AMSa}{"36}
      \DeclareMathSymbol{\geqslant}{3}{AMSa}{"3E}

      \let\leq=\leqslant \let\leq=\leqslant
      \let\geq=\geqslant \let\geq=\geqslant
    \fi
  \fi
\fi 

\ifCUPmtlplainloaded \else
  \ifAMStwofonts \else 
    \def\upi{\pi}
    \def\umu{\mu}
    \def\upartial{\partial}
  \fi
\fi

\title[Chemical Abundances in UV--Selected Galaxies]{Chemical abundances
in an UV--selected sample of galaxies}
\author[T. Contini et al.]
       {Thierry~Contini,$^{1,2}$\thanks{e-mail: {\tt contini@ast.obs-mip.fr}} Marie~A.~Treyer,$^3$ Mark~Sullivan$^4$ \& Richard~S.~Ellis$^5$\\
       $^1$ Observatoire Astronomique de Strasbourg, 11 rue de
       l'Universit{\'e}, F-67000 Strasbourg, France \\
       $^2$ Observatoire Midi-Pyr\'en\'ees, Laboratoire d'Astrophysique (UMR 5572), 14 avenue E. Belin, F-31400 Toulouse, France \\
       $^3$ Laboratoire d'Astrophysique de Marseille, Traverse du Siphon, F-13376 Marseille, France \\
       $^4$ Institute of Astronomy, Madingley Road, Cambridge, CB3 OHA, UK \\
       $^5$ California Institute of Technology, Pasadena, CA 91125, USA }
\date{Accepted ?.
      Received ?;
      in original form ?}

\pubyear{2001}

\begin{document}

\maketitle

\label{firstpage}

\begin{abstract}
  
  We discuss the chemical properties of a sample of UV-selected
  intermediate-redshift ($0 \la z \la 0.4$) galaxies in the context of
  their physical nature and star formation history. This work
  represents an extension of our previous studies of the rest-frame UV
  luminosity function (Treyer et al. 1998) and the star formation
  properties of the same sample (Sullivan et al. 2000, 2001).  We revisit
  the optical spectra of these galaxies and perform further
  emission-line measurements restricting the analysis to those spectra 
  with the full set of emission lines required to
  derive chemical abundances. Our final sample consists of 68 galaxies
  with heavy element abundance ratios and both UV and CCD $B$-band
  photometry.  Diagnostics based on emission-line ratios show that
  {\em all} but one of the galaxies in our sample are powered by hot,
  young stars rather than by an AGN. Oxygen-to-hydrogen (O/H) and
  nitrogen-to-oxygen (N/O) abundance ratios are compared to those of
  various local and intermediate-redshift samples. Our UV-selected
  galaxies span a wide range of oxygen abundances, from $\sim$ 0.1 to
  1 \zsun, intermediate between low-mass \hii\ galaxies and massive
  starburst nuclei. For a given oxygen abundance, most have strikingly
  low N/O values.  Moreover, UV-selected and \hii\ galaxies
  systematically deviate from the usual metallicity-luminosity
  relation in the sense of being more luminous by 2-3 magnitudes.
  Adopting the ``delayed-release'' chemical evolution model, we
  propose our UV-selected sources are observed at a special stage in
  their evolution, following a powerful starburst which enriched their
  ISM in oxygen and temporarily lowered their mass-to-light ratios. We
  discuss briefly the implications of our conclusions on the nature of
  similarly-selected high-redshift galaxies.

\end{abstract}


\section{Introduction}

Considerable progress has been made recently in understanding the
history of galaxy formation and evolution, following the analyses of
deep surveys, such as the {\it Hubble Deep Field} (Williams et al.
1996), the {\it Canada-France Redshift Survey} (Lilly et al. 1995) and
the population of {\it Lyman-break galaxies (LBGs)} (Steidel et al.
1996, 1999). A popular viewpoint is one that postulates that the bulk
of cosmic star formation occurred at redshifts between $\sim$ 1 and 2
(e.g., Madau, Pozetti \& Dickinson 1998), in which case the higher
redshift LBG population represents an early phase of galaxy formation.

However, the physical nature of the high redshift star-forming
galaxies remains unclear, and this hinders our understanding of how
they connect with present-day massive galaxies.  In particular, it
remains unclear to what extent the detectable (i.e.  bright and
star-forming) high-$z$ galaxies are representative of the entire
galaxy population that exists in the distant universe. The large
number of starburst-like objects detected questions the classical
``standard'' scenarios for the evolution of Hubble sequence galaxies,
which involve slowly decreasing, regular star formation histories.

Characterizing the {\it recurrence} of starbursts in galaxies is one
way of addressing the above issues. Do galaxies experience only one or
two major bursts of star formation during their life, or is the
starburst phenomenon a repetitive one that could mimic continuous star
formation on cosmological time-scales?
 
This question has been addressed recently by Kauffmann, Charlot \&
Balogh (2001), who explored numerical models of galaxy evolution in
which star formation occurs in two modes: a low-efficiency continuous
mode, and a high-efficiency mode triggered by interactions with a
satellite.  With these assumptions, the star formation history of
low-mass galaxies is characterized by intermittent bursts of star
formation separated by quiescent periods lasting several Gyrs, whereas
massive galaxies are perturbed on time-scales of several hundred Myrs
and thus have fluctuating but relatively continuous star formation
histories. In these models, merger rates are specified using the
predictions of hierarchical galaxy formation models (e.g. Kauffmann \&
Charlot 1998; Cole et al. 2000; Somerville, Primack \& Faber 2001).

Examining the chemical evolution and physical nature of star-forming
galaxies over a range of redshifts will shed light on this issue.
Emission lines from \hii\ regions have long been the primary means of
chemical diagnosis in local galaxies, but this method has only
recently been applied to galaxies at cosmological distances following
the advent of infrared spectrographs on 8 to 10-m class telescopes
(e.g. Steidel et al. 1996; Kobulnicky \& Zaritsky 1999; Kobulnicky \&
Koo 2000; Carollo \& Lilly 2001; Hammer et al. 2001; Pettini et al.
1998, 2001) .

The aim of this paper is to derive the chemical properties of a local
to intermediate-redshift ($0 \la z \la 0.4$) UV-selected sample of
galaxies as a further probe of their physical nature and star
formation history. A key finding from earlier papers in this series is
the possibility that the star formation in a large fraction of these
objects is intermittent, contributing significantly to the local UV
luminosity density (Sullivan et al. 2000, 2001).  The
manner in which this links to the chemical history may, when used
similarly with high redshift sources, give insight into the nature and
evolution of star-forming galaxies over a range of redshift.

Work on the nature of distant and young galaxies, like the LBGs,
greatly benefits from studies of their analogues in the nearby universe.
Early works revealed evidence that LBGs resemble local UV-bright
starburst galaxies in many respects (e.g. Heckman et al.  1998;
Meurer, Heckman \& Calzetti 1999; Papovich, Dickinson \& Ferguson
2001).  Although physically larger and more luminous, they share
similar star formation rates (SFR) per unit area (Meurer et al. 1997),
similar rest-frame UV-to-optical spectral energy distributions (Conti,
Leitherer \& Vacca 1996; Lowenthal et al. 1997; Pettini et al. 1998,
2001; Papovich, Dickinson \& Ferguson 2001), as well as similar 
interstellar medium dynamical states (Franx et al. 1997; Kunth
et al. 1998; Pettini et al. 1998, 2001).

A plan of the paper follows. In Section~\ref{uvsample} we give a brief
summary of the overall properties of our original UV-selected sample of
galaxies. In Section~\ref{elprop} we present new measurements of
emission lines and define a subsample of 68 UV-selected galaxies for
which heavy element abundance measurements are possible.  The
locus of the UV-selected galaxies in standard diagnostic diagrams is
investigated (Sect.~\ref{diag}) to determine their main source of
ionization. We search for the spectral signatures of Wolf-Rayet stars
(Sect.~\ref{wr}).  Empirical methods, based on strong emission-line
ratios, are presented in Section~\ref{chemabund} to derive the O/H and
N/O abundance ratios of the sample galaxies.  In Section~\ref{analysis}
we study how our sample compares with two fundamental scaling
relations: N/O versus O/H, and the metallicity-luminosity relation. We
summarize our principal conclusions in Section~\ref{conclu}. Throughout
this paper, all calculations assume an $\Omega = 1$ and $\rm H_0=100$
\kmsmpc\ cosmology.

\section[]{The parent UV-selected galaxy sample}
\label{uvsample}

A detailed description of the parent sample from which the analyses in
this paper are based can be found in Sullivan et al. (2000, hereafter 
S2000).  Briefly, the fields were
first imaged in the UV using the balloon-borne FOCA telescope, a 40cm
Cassegrain mounted on a stratospheric gondola, stabilised to within a
radius of 2\arcsec\ rms (see Milliard et al.  1992 for a full
description of the experiment).  The spectral response of the UV
filter approximates a Gaussian centred at 2015~{\AA}, FWHM 188~{\AA}.
The camera was operated in two modes -- FOCA 1000 (f/2.56, 2.3\degr)
and FOCA 1500 (f/3.85, 1.55\degr) -- with a large field-of-view well
suited to survey work. The limiting depth of the exposures we used is
\muv\ $=18.5$, which, for a late-type galaxy, corresponds to \mb\ 
$=20-21.5$.

The first field, Selected Area 57 (SA57), is centered on
$\rmn{RA}=13^{\rmn{h}}03^{\rmn{m}}53^{\rmn{s}}$, $\rmn{Dec.}=+29\degr
20\arcmin 30\arcsec$ (1950) and contains the outer regions of the Coma
cluster. The second field is centered on $\rmn{RA}=11^{\rmn{h}}
42^{\rmn{m}} 46^{\rmn{s}}$, $\rmn{Dec.}=+20\degr 10\arcmin 03\arcsec$,
and contains the Abell 1367 cluster. Both fields were imaged in both
FOCA modes, thus ensuring the most reliable UV photometry. As the
astrometric accuracy of FOCA ($\simeq$ 3\arcsec\ rms for FOCA 1500) is
insufficient for generating a spectroscopic target list, the FOCA
catalogues were matched with APM scans of the POSS optical plates.
When more than one possible optical counterpart was found on the POSS
plates within the search radius, the nearest optical counterpart to the
UV detection was selected.  Around 10\% of the UV detections have no
obvious counterpart at all on the APM plates, indicating that either
some of these detections are spurious, or that the counterpart lies at
a fainter $B$ magnitude than the limiting magnitude of the POSS plates
(\mb\ $\simeq 21$).

In order to improve the photometric coverage of our survey fields, an
extensive, high quality multi-colour survey is currently underway. In this
present work, we use new $B$-band CCD photometry obtained with the CFH12K 
camera on the
Canada-France-Hawaii Telescope (CFHT) in place of the APM magnitudes
used in previous studies (Treyer et al.  1998; S2000).  Further
details of this photometry and is relationship to that used earlier
will be given in a forthcoming paper (Sullivan et al., in preparation) .

Spectroscopic follow-up of the UV selected sources was conducted on
the two FOCA fields using two multi-fiber spectrographs -- Hydra on
the 3.5-m WIYN telescope ($\lambda\lambda$ 3500--6600~{\AA}), and
WYFFOS on the 4.2-m William Herschel Telescope (WHT) ($\lambda\lambda$
3500--9000~{\AA}) (Treyer et al. 1998, S2000). The latter
configuration allows the observation of \halpha\ emission lines to a
redshift of $z \sim 0.4$. Only the flux calibrated data from WYFFOS are 
used in the present paper. Details of all observing runs and the
breakdown of spectroscopic objects can be found in S2000, tables 1 and
2 respectively.

Spectra were analysed using the \textsc{iraf} facility \textsc{splot}
and the \textsc{figaro} package \textsc{gauss}.  Redshifts were
measured by visual inspection, and the equivalent widths (EWs) and
fluxes of \oii, \oiii, \hbeta\ and \halpha\ determined using both
spectral analysis programs. Though the spectral resolution ($\sim$
10~{\AA}) is good enough to resolve the separate \oiii\ lines, in many
cases the \halpha\ line was blended with the nearby \Nii\ lines, so a
deblending routine was run from within \textsc{splot} to allow
determination of the fluxes of these individual lines. 
We did not find any systematic error in the measurement of individual lines 
which could be due to the deblending procedure. 
Extinction and
stellar absorption corrections were applied using the measured Balmer
lines.  A full description of these procedures can be found in S2000.
Typical examples of galaxy spectra are shown in Treyer et al. (1998,
their fig. 1).

\section[]{
Emission-line properties
}
\label{elprop}

\begin{figure}
\vspace{-4.2cm}
 \epsfig{figure=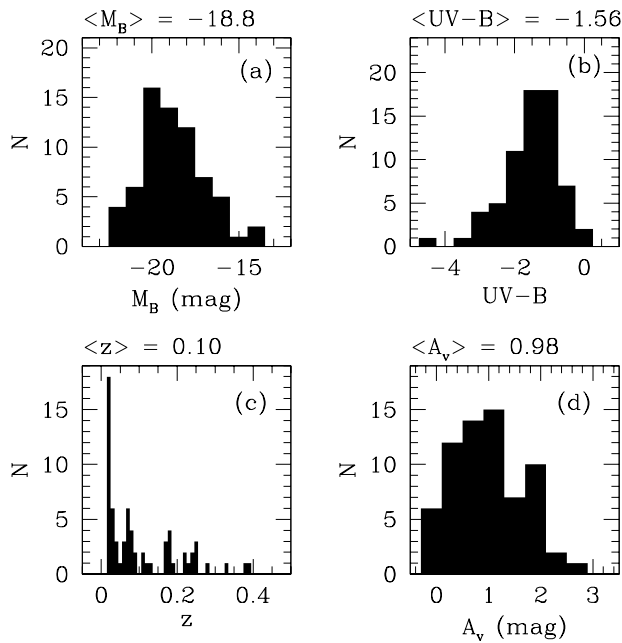,width=130mm}
 \caption{Distributions of a) the $B$-band absolute magnitudes,
b) the dereddened $UV - B$ colours, c) the redshifts, and d) the extinction
coefficients of the UV-selected galaxy sample with
measured chemical abundances. Mean value of the distributions is
indicated on top of each panel. The peak at $z \sim 0.02$ in the redshift
distribution arises from the Coma and Abell 1367 galaxy clusters.}
 \label{histomag}
\end{figure}

The analysis of S2000 was essentially concerned with the star
formation properties of the UV-selected sample described above, using
UV fluxes, Balmer and \oii\ emission lines.  Emission-line properties
were only briefly discussed with the aim of estimating a possible AGN
contamination or metallicity effect in order to explain the extreme
($UV-B$) colours found in $\sim 17$\% of the galaxies.  Based on
\oii/\hbeta\ and \oiii/\hbeta\ line ratios, S2000 found that a
fraction of the UV galaxies may indeed qualify as AGNs (their fig.
11). However, the \oii/\hbeta\ diagnostic used in that study is very
sensitive to dust extinction corrections, and it is therefore
preferable to use the \nii/\halpha\ and \sii/\halpha\ line ratios
which are much less sensitive to extinction (see sect.~\ref{diag}).
S2000 also searched for possible extremely metal-poor objects in the
UV-selected sample which could account for the extreme ($UV-B$)
colours. However oxygen abundances were estimated for only 35 field
galaxies in the SA57, and no conclusive results were found.

The goal of this present work is to examine in a more systematic
manner the emission-line properties of the UV-selected galaxies. In
order to make diagnostic diagrams and estimate chemical abundances for
a larger fraction of the sample, we revisit the sample of S2000 and,
where an emission line measurement is missing, attempt a new line
measurement. We then restrict the S2000 sample to all the galaxy spectra 
which posses the full set of emission lines (i.e. \oii, \hbeta, \oiii, 
\halpha\ and \nii) required to derive chemical abundances.  Our final
sample consists of 68 galaxies with heavy element abundance ratios, UV
and new CCD $B$-band photometry.

\subsection[]{
New measurements of emission lines
}
\label{newmes}

For consistency with the spectral measurements of S2000, the new
emission-line measurements were performed using the {\sc splot}
facility in {\sc iraf}.  In particular, we searched for \sii\ emission
lines in all of the 135 emission-line galaxies in the parent sample,
and could make a reliable measurement of these lines in 56 of the objects.
Additional measurements of \hbeta\ (8 objects), \oiii\ (15 objects),
and \nii\ (10 objects) could also be performed.  Reddening
corrections were applied to the new emission-line fluxes as in S2000.
The extinction coefficient \cbeta\ was estimated from the Balmer
decrement \halpha/\hbeta, assuming case B recombination with an
electron density of 100 \cmc\ and a temperature of $10^4$ K
(Osterbrock 1989), and using a standard interstellar extinction law
(Seaton 1979).  As in S2000, \halpha\ and \hbeta\ were also corrected
for stellar absorption (prior to computing \cbeta).

There are two major sources of uncertainty in the measured
emission-line ratios. The first arises from the limited signal/noise
ratio of the spectra. These errors are computed directly when using
the {\sc iraf} task {\sc splot} to derive the emission line
parameters. {\sc splot} provides 1-$\sigma$ errors based on estimates
of the noise in the individual spectra. The integration error
estimates are derived by error propagation assuming a Poisson
statistic model of the pixel sigmas, generated by measuring the noise
in the spectra on an individual basis and assuming that the linear
continuum has no error (see S2000).  Measurement errors were
propagated quadratically when deriving emission-line ratios and
abundance ratios (cf.  Sect.~\ref{chemabund}).

The second source of error arises from contamination of Balmer
emission lines by underlying stellar absorption lines. In most cases,
Balmer emission line equivalent widths were directly measured in the
galaxy spectra. When this was not possible, a constant value of 2
{\AA}\, corresponding to the average measured value, typical for
star-forming galaxies (McCall et al. 1985; Olofsson 1995a;
Gonz{\'a}lez-Delgado, Leitherer \& Heckman 1999), was applied. This
might be a poor approximation for galaxies with weak \hbeta\ emission
line. We thus propagated an uncertainty of $\pm 1$ {\AA}\ on Balmer
absorption line equivalent widths in all calculations involving Balmer
emission lines.  Finally, the error in the derived extinction
coefficient \cbeta\ was included when deriving dereddened
emission-line ratios.

Global properties of the sub-sample of 68 UV-selected galaxies are
shown in Fig.~1. Dust-corrected UV and $B$-band absolute magnitudes
are computed as in S2000 using Calzetti (1997) extinction curve. Note
that the dereddened ($UV-B$) colour and $B$-band absolute magnitude
distributions differ from those of the original sample (S2000) due to
the revised $B$-band photometry. The red tail of the ($UV-B$) colour
distribution has been significantly reduced whilst the blue component
-- which formed an interesting unresolved aspect of the discussion in
S2000 -- is largely unchanged. The revised optical colours means that
a a small fraction of the sample is now assigned a different spectral
type and $k$-correction. The effect on the absolute UV luminosities
are fairly modest however. The main impact of the revised CCD
photometry will be in estimating the spectral energy distributions and
consequent $k$-corrections; a topic we will discuss in the forthcoming
paper (Sullivan et al., in preparation).

\subsection[]{
Nature of the main ionizing source: starburst or AGN?}
\label{diag}

\begin{figure*}
 \epsfig{figure=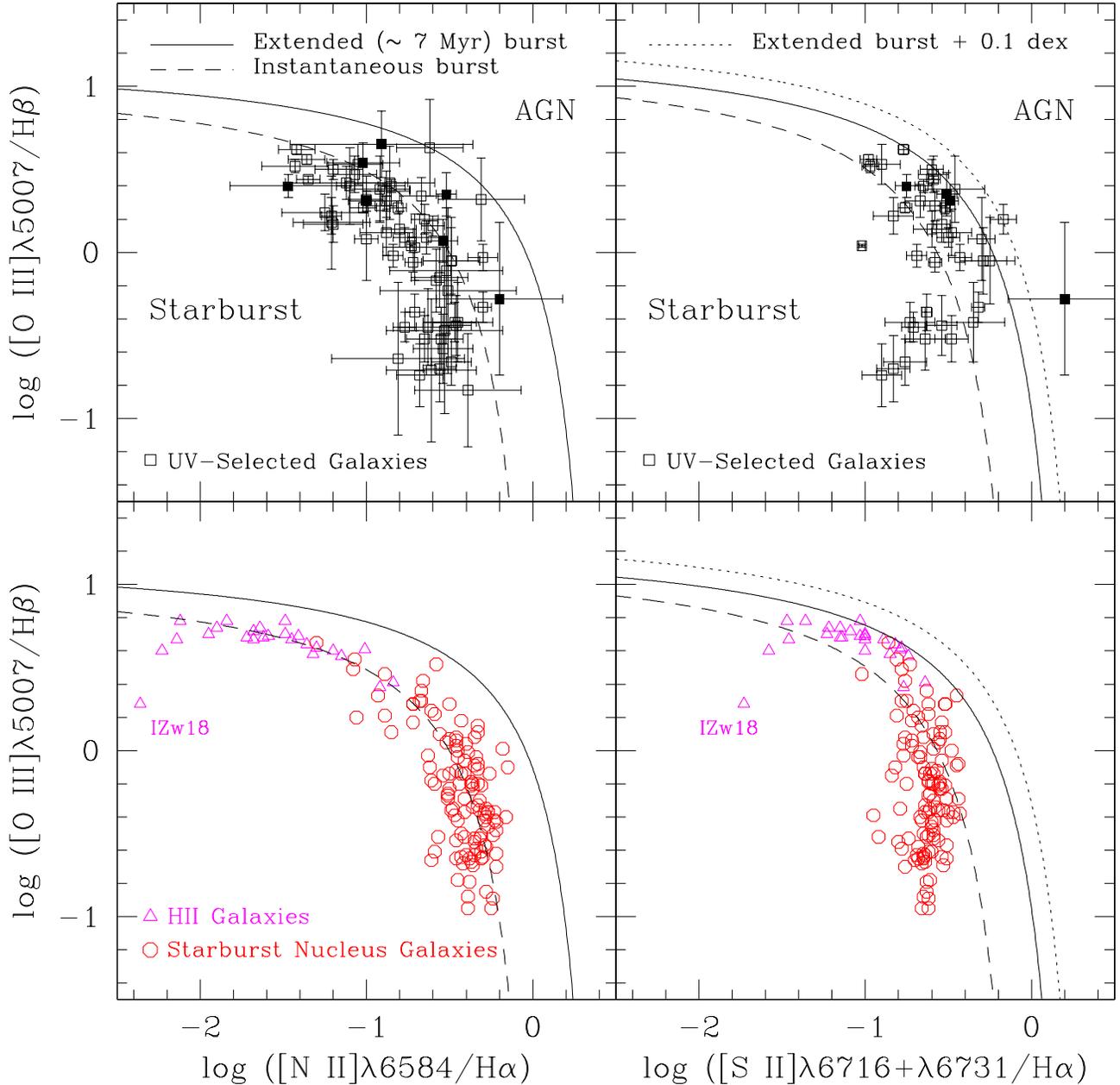,width=180mm}
 \vspace{-1cm}
 \caption{Standard diagnostic diagrams used to classify
   narrow emission-lines spectra. {\it Top panels}. The UV-selected
   galaxies are shown with their error bars; filled squares are double
   optical counterpart cases (see text for details).  The curves are
   theoretical boundaries between starburst galaxies and AGNs,
   assuming an ``extended'' ($\sim 7$ Myr) burst of star formation
   (solid line; Kewley et al. 2001) or an ``instantaneous'' starburst
   (dashed line; Dopita et al. 2000). The dotted line in the
   \sii/\halpha\ vs.  \oiiib/\hbeta\ diagram is the upper limit of the
   starburst model (see text for details). These diagrams show that
   {\em all} the UV-selected galaxies, but one uncertain case which
   might qualify as a LINER, are powered by hot and young stars rather
   than by an AGN. {\it Bottom panels}. Samples of nearby star-forming
   galaxies are shown for comparison: Starburst Nucleus Galaxies
   (circles) selected in the optical (Contini et al. 1998) or in the
   far-infrared (Veilleux et al. 1995); and \hii\ galaxies (triangles;
   see text for references).  The location of IZw18, the most
   metal-poor galaxy known so far, is indicated.  }
\label{diagnostic}
\end{figure*}

Emission-line diagnostic diagrams can provide a reliable classification
of narrow emission-line spectra according to the main source of
ionization (hot stars, AGN, or shocks) responsible for these lines.  
We examine the \oiiib/\hbeta\ versus \nii/\halpha\ and
\sii/\halpha\ diagnostic diagrams to discriminate regions photoionized
by hot and young stars (i.e. \hii\ regions) from those photoionized by
a harder radiation field, such as that of an AGN or a LINER. Such
emission-line ratios are much less sensitive to dust extinction than
\oii/\hbeta, the line ratio previously used in S2000 for this purpose.

The location of the UV-selected galaxies in these diagnostic diagrams
is shown in Fig.~\ref{diagnostic}, using dereddened emission-line
ratios with their related uncertainties listed in
Table~\ref{abundtab}.  In each diagram, the theoretical curves
(discussed below) separate star-forming regions (lower left) where the
gas is assumed to be ionized by young stars, from AGNs (upper right)
where the main ionizing source is thought to be an accretion  disk
around a black hole.  A further distinction among AGN-like objects can
be made between high-excitation (\oiiib/\hbeta\ $>3$) Seyfert 2
galaxies and low-excitation (\oiiib/\hbeta\ $<3$) LINERs. We did not
use the original criteria for defining a LINER (Heckman 1980) because
the measurement of \oi\ emission lines was generally not possible due
to the insufficient S/N of the spectra. 
For the same reason, we did not use the
\oi/\halpha\ ratio to distinguish between the different sources of
ionization.

Two comparative star formation models are shown in
Fig.~\ref{diagnostic}: the solid line represents an ``extended'' burst
(i.e. one extending for $\sim 7$ Myr, Kewley et al. 2001) and the
dashed line represents an ``instantaneous'' burst (Dopita et al.
2000).  In both cases, the parameters were chosen to span a realistic
range of metallicities ($Z=0.1-3.0$ \zsun) and ionization parameters
($-3.5\leq \log U\leq -2.0)$. These new theoretical boundaries may
provide a more objective classification between \hii\ regions and the
various classes of narrow-line regions associated with AGNs than the
classical semi-empirical boundaries defined by Veilleux \& Osterbrock
(1987).

Clearly, these model predictions have associated uncertainties.  These
include the assumed chemical abundances and the depletion factors, the
slope of the IMF,  the stellar evolutionary tracks and the model
atmospheres. The dotted line in Fig.~\ref{diagnostic} (right panel)
gives an indication of these uncertainties in the \sii/\halpha\ vs.
\oiiib/\hbeta\ diagram and represents an upper limit to the theoretical
boundary between starbursts and AGNs, corresponding to the
``extended burst'' model $+0.1$ dex (see
Kewley et al. 2001 for details).

Figure~\ref{diagnostic} shows that nearly {\em all} the UV galaxies lie
below and to the left of the theoretical starburst line. This is
especially clear from the \nii/\halpha\ vs. \oiiib/\hbeta\ diagram,
where all the galaxies form a well-defined sequence of
\hii\ region-like objects. The distinction is more ambiguous in the
\sii/\halpha\ vs. \oiiib/\hbeta\ diagram. On this plot, a significant
fraction of the galaxies are borderline AGN candidates, although they
still lie within the error bars of the ``extended burst'' star
formation model.

One galaxy (ID \#\,24 in Tables~\ref{abundtab} and \ref{magtab}) with
a relatively high \sii/\halpha\ ratio lies on the right-hand side of
the diagram, and could be considered a LINER-type object. However, this
galaxy is one of two possible optical counterparts to the UV source.
Moreover, the redshifted \sii\ lines at $z$=0.28 are highly
contaminated by bright sky lines, giving a larger uncertainty in this
diagnostic than for the bulk of the other sources.

It is instructive to compare the location of the UV-selected galaxies
in these diagnostic diagrams with other published samples of nearby
star-forming galaxies selected in optical and far-infrared bands.
Emission-line ratios for \hii\ galaxies have been published by Izotov
\& Thuan (1998) and Izotov, Thuan \& Lipovetsky (1994, 1997), and for
starburst nucleus galaxies (SBNGs) by Contini et al. (1998) and
Veilleux et al. (1995). Both form a well-defined sequence in our two
diagnostic diagrams. \hii\ galaxies are characterized by a high
excitation level (\oiiib/\hbeta\ $\ga 1.5$) and low \nii/\halpha\ and
\sii/\halpha\ emission-line ratios, whereas SBNGs have lower
excitation levels (\oiiib/\hbeta\ $\la 1$) and higher \nii/\halpha\ 
and \sii/\halpha\ emission-line ratios.

These nearby star-forming galaxies are very well reproduced by the
``instantaneous'' star formation model in the \nii/\halpha\ vs.
\oiiib/\hbeta\ plot, but this is {\em not} the case for the
\sii/\halpha\ vs. \oiiib/\hbeta\ diagram. The behavior of SBNGs
in this respect is particularly interesting. As the excitation level
decreases, the \nii/\halpha\ ratio increases whereas the \sii/\halpha\
stay roughly constant. Note also that the star formation models
overpredict (by $\sim 0.3$ dex) the \sii/\halpha\ line ratio for SBNGs with
the lowest excitation levels.

Figure~\ref{diagnostic} shows that the major fraction of UV galaxies
have excitation levels which are similar to SBNGs whereas extreme
\hii\ galaxies with very low \nii/\halpha\ and \sii/\halpha\ ratios
are not present in the UV-selected sample. In detail, however, the
sequence defined in the \nii/\halpha\ vs. \oiiib/\hbeta\ diagram is
somewhat different from the one traced by SBNGs.  This is especially
true for the low-excitation (\oiiib/\hbeta\ $\la 1$) UV objects, for
which \nii/\halpha\ is $\sim 0.1-0.2$ dex lower than for SBNGs.  This
could be due to chemical enrichment effects. Indeed it has been shown
that SBNGs possess a slight overabundance of nitrogen compared to
\hii\ regions with comparable metallicity (Coziol et al. 1999,
Consid{\`e}re et al. 2000).  This point will be further discussed in
section~\ref{novsohrel}.

The main conclusion, however, is that the emission-line spectra of
{\em all} the UV galaxies in our sample, but one dubious case, are
powered by hot and young stars rather than by an AGN.  We are thus
able to derive the chemical properties of the ionized gas in these
galaxies using the standard recipes applied for star-forming galaxies.

\subsection[]{
Search for Wolf-Rayet stars
}
\label{wr}

\begin{figure}
 \epsfig{figure=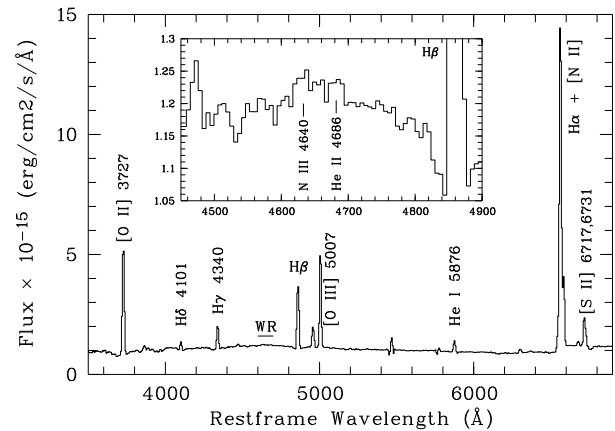,angle=-90,width=85mm}
 \caption{Restframe optical spectrum of an UV-selected galaxy at
$z=0.0205$ (ID \# 56 in Tables~\ref{abundtab} and \ref{magtab}). 
Broad \heii\ and \niii\ emission lines typical
of WR stars are detected at the 3$\sigma$ level in the blue
part of the spectrum.
}
 \label{wrgal}
\end{figure}

One of the most puzzling features of the S2000 sample was the presence
of a population of galaxies with UV-optical colours too blue ($UV-B
\la -2$) for consistency with standard population synthesis models of
starburst galaxies. This UV excess was observed in 17\% of the (single
optical counterpart) S2000 galaxies. One possible explanation of this
effect was posited by Brown et al. (2000), who argued that UV emission
lines from hot Wolf-Rayet (WR) stars could result in the UV excess
observed.  The present sample contains 23 such sources of which 8 have
double optical counterparts and therefore may represent possible
misidentifications of the underlying UV detection (see
Sect.~\ref{uvsample}). As described earlier (sect.~\ref{newmes}), the
revised $B$-band optical photometry does not change the blue tail of
the $UV-B$ colour distribution.  This confirms that the observed UV
excess is not simply due to large uncertainties in the photometry.

As well as investigating Brown et al's hypothesis, there is a further
motivation in examining our spectra for WR features.  WR stars, which
are the direct descendants of the most massive O stars, are predicted
in large numbers in young star-forming regions by population synthesis
models (e.g., Meynet 1995; Schaerer \& Vacca 1998; Leitherer et al.
1999).  They have been extensively used to quantify starburst
parameters (age, duration and IMF) and to provide constraints on both
population synthesis and massive star evolution models (e.g., Vacca \&
Conti 1992; Contini et al. 1995; Schaerer, Contini \& Kunth 1999;
Guseva et al. 2000; Schaerer et al. 2000).  Optical integrated spectra
of the so-called ``WR galaxies'' show direct signatures from WR stars,
most commonly a broad \heii\ feature originating in the stellar winds
of these stars. Many objects were recently found to harbour additional
features from WR stars in their spectra. For example, the broad
\niii\, \civ\ and \ciii\ emission lines, among the strongest optical
lines in WN and WC stars, are more and more often detected.

Although weaker and thus requiring higher S/N spectra, lines
originating from WC stars (i.e. \civ\ and \ciii), representing more
evolved phases than WN stars, provide complementary information on the
massive star content in galaxies (e.g., Schaerer, Contini \& Kunth
1999; Guseva et al. 2000; Schaerer et al. 2000). Since the initial
compilation of Conti (1991) listing 37 objects, the number of known WR
galaxies reaches $\sim 140$ in the last catalog of Schaerer, Contini \&
Pindao (1999). Six new WR galaxies have recently been reported by 
Popescu \& Hopp (2000).

We find that only one galaxy at $z=0.0205$ (ID \#\,56 in
Tables~\ref{abundtab} and \ref{magtab}) clearly shows (3$\sigma$)
broad \heii\ and \niii\ emission lines (Fig.~\ref{wrgal}). There are
four other WR galaxy ``candidates'' with marginal detections ($\sim
1\sigma$).  However, none of these five objects shows a UV excess.
Their ($UV-B$) colours lie between $-1.2$ and $0.25$, redder than the
sample mean value of $-1.6$ (see Fig.~\ref{histomag}b). Moreover, no
WR spectral features could be identified in the spectrum of galaxies
with $UV-B \la -2$.  We conclude that the presence of hot WR stars is
not the best explanation to account for the UV excess in these
galaxies, as suggested by Brown et al. (2000).
Conceivably some fraction of the UV light of these galaxies
could arise from a non--thermal source (i.e. low-luminosity AGN) 
or, alternatively, 
the UV emission may somehow be independent of the optical radiation. 
High-resolution images capable of resolving the distribution of 
intense UV radiation will help to address such possibilities, 
and/or to identify the shortcomings of current stellar population 
synthesis models. 

Using emission line luminosities, we can roughly estimate the massive
stellar population in the WR galaxy presented in Figure~\ref{wrgal}.
 As the strength of the \niii\ emission line
($EW = 1.2$ {\AA}) is comparable to that of the \heii\ line ($EW = 0.8$
{\AA}), most likely the dominant subtype of WN stars is late-type WN
(WNL). Not much can be said about WC-type stars which are more
frequently seen in WR galaxies (see Schaerer, Contini \& Kunth 1999;
Guseva et al. 2000).  The strongest line, \civ\, produced by this
subtype is not seen in the spectrum but this may be due to the presence
of a strong sky line in the same wavelength range.

The number of WNL stars can be estimated from the observed \heii\ line
luminosity and adopting the average observed luminosity of
WNL stars in the \heii\ line ($1.6 \times 10^{36}$ \ergs; Schaerer \&
Vacca 1998). The dereddened flux of the \heii\ line is $5.6 \times 
10^{-15}$ \ergscm\ corresponding, at $z=0.0205$, to a luminosity 
of $L_{\rm He\,II} \sim 2.6\times 10^{39}$ \ergs. 
Roughly, we find a value of about 1600 WNL stars.  The
number of O stars can be derived from the dereddened
\hbeta\ emission-line luminosity, which is equal to $\sim 1.2 \times
10^{41}$ \ergs.  We follow the procedure outlined in Schaerer, Contini
\& Kunth (1999) which takes into account the ionizing photon
contribution from WR stars. The resulting number of O stars is between
$5\,000$ and $20\,000$, depending on the IMF parameters. This gives a
WR/O star number ratio in the range of $0.06-0.25$, similar to what is
found in typical WR galaxies (e.g. Schaerer, Contini \& Kunth 1999).
This ratio is systematically higher than the predictions for constant
star formation at the appropriate metallicity ($Z \sim 0.8$ \zsun;
Maeder \& Meynet 1994), but within the range of instantaneous burst
models with different IMF slopes (Schaerer \& Vacca 1998).

\begin{figure}
 \epsfig{figure=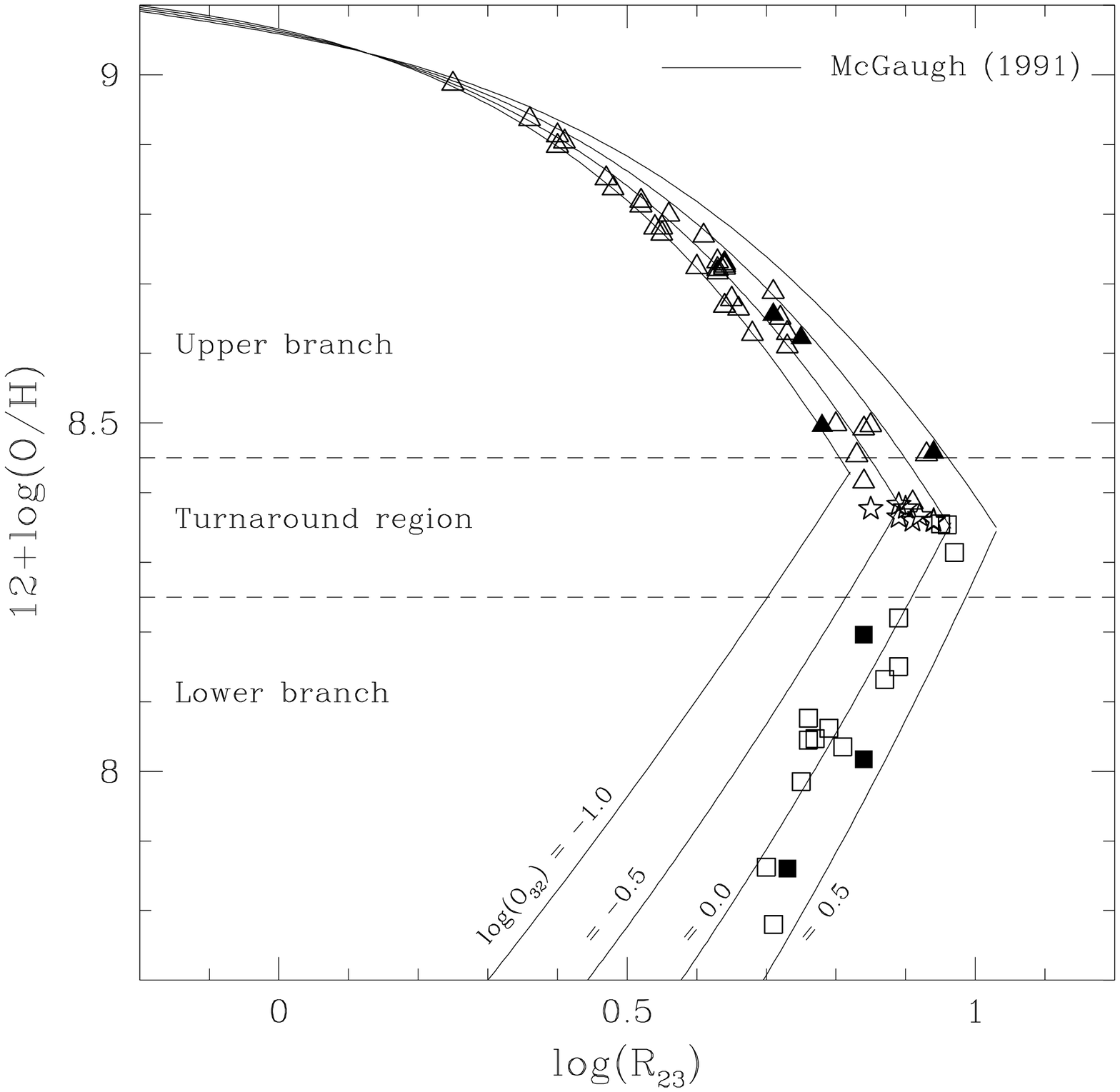,width=85mm}
 \caption{Calibration of oxygen abundance \doh\ as a function
of the strong-line ratio \rr23 $\equiv$ (\oii+\oiii)/\hbeta.
The solid lines show the effect of varying the ionization parameter
expressed in terms of the observable line ratio \oo32 $\equiv$ \oiii/\oii\ 
(McGaugh 1991). The relationship is degenerate and therefore requires
some a priori knowledge of a galaxy's metallicity to
determine its position either on the {\em upper} (\doh\ $> 8.45$)
or on the {\em lower} (\doh\ $< 8.25$) branch of the curve (see text).
The UV-selected galaxies are plotted with
the following symbols: {\it squares}, the galaxy lies on the
{\em lower} branch; {\it triangles}, the galaxy lies on the
{\em upper} branch; {\it stars}, the galaxy falls in the turnaround
region where the \nii/\halpha\
and \nii/\oii\ diagnostics are inconclusive (see Fig.~\ref{method}).
Filled symbols are UV-selected galaxies with two optical counterparts.
}
 \label{calib}
\end{figure}

\section[]{Chemical abundances}
\label{chemabund}

\subsection[]{Oxygen abundance}

Emission lines are the primary source of information regarding chemical
abundances within \hii\ regions.  Because nebular cooling occurs
principally through the escape of photons generated in spontaneous
de-excitation of metallic ions (e.g. oxygen, nitrogen, and sulphur),
the strength of the emission lines of these different species is an
indicator of the electronic temperature.  For oxygen, the cooling in
the nebulae occurs primarily either via fine-structure lines in the
far-infrared (52 and 88 \micron) when the electron temperature is low
or via forbidden lines in the optical (\oii, \oiiia, and \oiiib) when
the electron temperature is high.  Since the temperature is high when
there is insufficient metal line cooling, strong optical
lines imply low metallicity.

The ``direct'' method for determining chemical compositions requires
the electron temperature and the density of the emitting gas  (e.g.,
Osterbrock 1989). Unfortunately, a direct heavy element abundance
determination, based on measurements of the electron temperature and
density, cannot be obtained for our sample.  The \oiiic\ auroral line,
which is the most commonly applied temperature indicator in
extragalactic \hii\ regions, is typically very weak and rapidly
decreases in strength with increasing abundance; it is expected to be
of order $10^2 - 10^3$ times fainter than the \oiiib\ line.

Given the absence of reliable \oiiic\ detections in our faint spectra,
alternative methods for deriving nebular abundances that rely on
observations of the bright lines alone must be employed.  Empirical
methods to derive the oxygen abundance exploit the relationship between
O/H and the intensity of the strong lines via the parameter
\rr23\ $\equiv$ (\oii+\oiii)/\hbeta\ (see Fig.~\ref{calib}).

Many authors have developed techniques for converting \rr23\ into
oxygen abundance, both for metal-poor (Pagel, Edmunds, \& Smith
1990; Skillman 1989; Pilyugin 2000) and metal-rich (Pagel et al.
1979; Edmunds \& Pagel 1984, McCall et al. 1985; Dopita \& Evans
1986; Pilyugin 2001) regimes. In the most metal-rich \hii\
regions, \rr23\ is minimal because metals permit efficient
cooling, reducing the electronic temperature and the level of
collisional excitation. On the {\em upper}, metal-rich branch of
the relationship, \rr23\ increases as metallicity decreases via
reduced cooling, elevated electronic temperatures, and a higher
degree of collisional excitation. However, the relation between
\rr23\ and O/H becomes degenerate below \doh\ $\sim$ 8.4 ($Z \sim
0.3$ \zsun). As metallicity decreases below \doh\ $\sim 8.2$,
\rr23\ decreases once again. On this {\em lower}, metal-poor
branch, although the reduced metal abundance further inhibits
cooling and raises the electron temperature, the intensity the
\oii\ and \oiii\ lines drops because of the greatly reduced oxygen
abundance in the ionized gas. In this regime, the ionization
parameter also becomes important (McGaugh 1991).

The different ionization parameters may lead to two very different
oxygen abundances for a single value of \rr23 as illustrated in
Figure~\ref{calib}. We represent the approximate ionization
parameter in terms of the easily observable line ratio \oo32
$\equiv$ \oiii/\oii. The solid lines show the oxygen abundance as
a function of \rr23\ for log(\oo32) = $-$1, $-$0.5, 0 and 0.5
corresponding, very roughly, to ionization parameters between
$\sim$ 1 and $10^{-2}$. Figure~\ref{calib} is thus useful for
finding the oxygen abundance of nebulae when the electron
temperature cannot be measured directly. The typical uncertainty
is $\pm 0.15$ dex, although it is larger ($\pm 0.25$ dex) in the
turnaround region near \doh\ $\sim 8.4$. This dispersion
represents the uncertainties in the calibration which is based on
photoionization models and observed \hii\ regions. However, the
most significant uncertainty involves deciding whether an object
lies on the {\em upper}, metal-rich branch, or on the {\em lower},
metal-poor branch of the curve.

\begin{figure}
\vspace{-0.3cm}
 \epsfig{figure=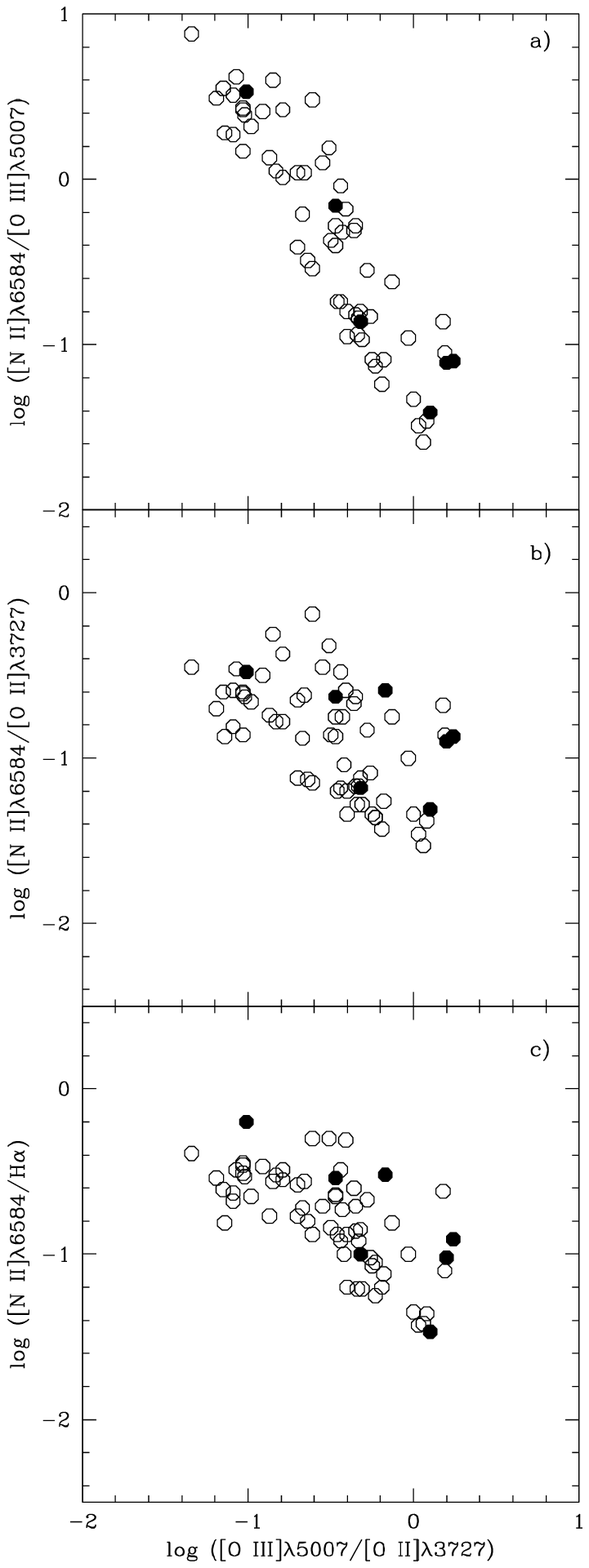,width=210mm}
 \caption{Various abundance indicators as a function of the ionization
indicator \oiiib/\oii\ for the UV-selected galaxy sample.
Filled circles are double counterpart cases as in previous plots.
The \nii/\oiiib\ line ratio ({\it a}), first proposed
by Alloin et al. (1979) as an abundance indicator, is very sensitive
to the ionization parameter. The \nii/\oii\ and \nii/\halpha\
line ratios ({\it b} and {\it c}) are much less sensitive to it and are
therefore better abundance indicators. We use them to
break the degeneracy in the O/H vs. \rr23\ relationship.
}
 \label{ioniz}
\end{figure}

Several methods have been proposed to break this degeneracy. The
abundance indicator \nii/\oiiib, was first proposed by Alloin et
al. (1979). \nii/\oiiib\ is usually lower than $\sim 10^{-2}$ for
galaxies on the upper metal-rich branch. This is because objects
which are considerably enriched in oxygen are generally more
nitrogen-rich as well, while the most metal-poor galaxies on the
lower branch of the \rr23\ relation have very weak \nii\ lines.
While this parameter varies monotically with abundance (e.g.,
Edmunds \& Pagel 1984), it is very sensitive to the ionization
parameter and is thus of limited use in the low-abundance regime,
where such effects are important (see Fig.~\ref{calib}). Moreover,
this ratio depends on the nucleosynthetic origin of nitrogen and
on the details of the galaxy star formation history. Metallicities
could thus be overestimated if nitrogen was enriched in the
nuclear region of starburst galaxies (e.g. Coziol et al. 1999;
Consid{\`e}re et al. 2000).

The \nii/\oii\ line ratio has also been used (e.g., McGaugh 1994;
van Zee et al. 1998). This line ratio varies monotically with
abundance, forms a narrow sequence over a large range of
metallicity (McCall et al. 1985), and is much less sensitive to
the ionization parameter. The division between upper and lower
branches occurs around \nii/\oii\ $\sim 0.1$. In general, \hii\
regions with \nii/\oii\ $< 0.1$ are believed to have low oxygen
abundances, while those with \nii/\oii\ $> 0.1$ are on the
high-metallicity branch. The \nii/\oii\ diagnostic is however
inconclusive in the turnaround region, that is for $-1.05 <$
log(\nii/\oii) $< -0.8$. Moreover, this line ratio suffers from
the same uncertainties as \nii/\oiiib\ regarding the
nuclesynthetic origin of nitrogen and is much more sensitive to
extinction corrections.

A third ratio, \nii/\halpha, has also been proposed (van Zee et
al. 1998). Like the previous two diagnostics, \nii/\halpha\
increases with increasing oxygen abundance. It is also much less
sensitive to the ionization parameter than \nii/\oiiib. The
division between upper and lower branches occurs around
\nii/\halpha\ $\sim 0.1$. In general, \hii\ regions with
\nii/\halpha\ $< 0.1$ are believed to have low oxygen abundances,
while those with \nii/\halpha\ $> 0.1$ are on the high-metallicity
branch. But as noted by van Zee et al. (1998), the \nii/\halpha\
ratio is only valid as a metallicity estimator for \doh\ $< 9.1$
and in the absence of additional excitation sources (low-intensity
AGN or shocks) which could increase it. Furthermore, with typical
errors of 0.2 dex or more, it is not a particularly accurate
abundance estimator. However it provides an additional diagnostic
to reduce the ambiguity in the O/H vs. \rr23\ relation.

Finally, Kobulnicky, Kennicutt \& Pizagno (1999) proposed to use
the galaxy luminosity to break the degeneracy. Because
galaxies of all morphological types {\em in the local universe}
follow a luminosity-metallicity relation (e.g., Skillman,
Kennicutt, \& Hodge 1989, Zaritsky, Kennicutt, \& Huchra 1994;
Coziol et al. 1998), objects more luminous than \mabs\ $\simeq
-18$ have metallicities larger than \doh\ $\simeq 8.3$, placing
them on the upper branch of the curve.  However, it has not yet
been established whether star-forming galaxies and objects at
earlier epochs follow the same relationship as local galaxies (see
Sect.~\ref{metlumrel}).

Figure~\ref{ioniz}a,b, and c show for our UV-selected sample the
three line ratios discussed above as a function of the oxygen
excitation ratio \oiiib/\oii, which can be used as a good
indicator of the ionization parameter (e.g., McGaugh 1991).
Clearly \nii/\oiiib\ is very sensitive to the ionization
parameter, and thus cannot be used as a reliable abundance
indicator (Fig.~\ref{ioniz}a). On the other hand, \nii/\halpha\
and \nii/\oii\ are virtually independent of the ionization
parameter (Figs.~\ref{ioniz}b, and c). We propose to use both in
order to discriminate between the {\em lower} and {\em upper}
branches in the O/H vs. \rr23\ relationship.

\begin{figure}
 \epsfig{figure=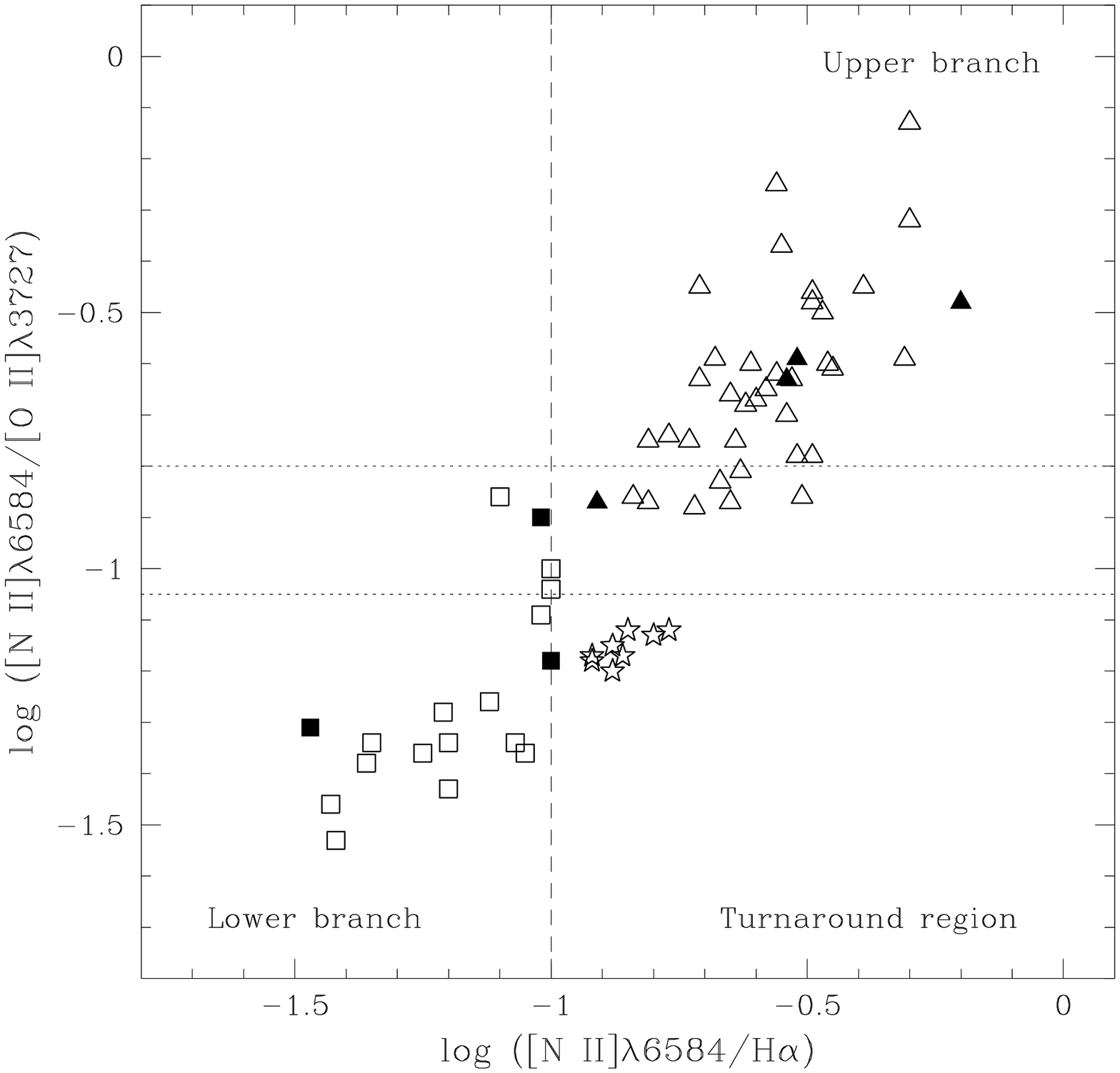,width=85mm}
 \caption{Emission-line ratios used to break the degeneracy
in the O/H vs. \rr23\ relationship. Symbols are as
in Fig.~\ref{calib}. A galaxy lies on the
{\em lower} branch (squares) if log(\nii/\halpha) $< -1$ and
log(\nii/\oii) $< -0.8$; on the {\em upper}
branch (triangles) if log(\nii/\halpha) $> -1$ and
log(\nii/\oii) $> -1.05$. A significant fraction (star symbols)
fall in the turnaround region of the O/H vs. \rr23\ relationship,
where the \nii/\halpha\ and \nii/\oii\ diagnostics are inconclusive.
}
 \label{method}
\end{figure}

The oxygen abundance can be determined using the calibrations of
McGaugh (1991). Analytic expressions are found in Kobulnicky,
Kennicutt \& Pizagno (1999), both for the metal-poor ({\em lower})
branch:
\begin{eqnarray}
12+\log({\rm O/H})=7.056+0.767x+0.602x^2 \cr
                   -y(0.29 +0.332x-0.331x^2),
\end{eqnarray}
and for the metal-rich ({\em upper}) branch:
\begin{eqnarray}
12+\log({\rm O/H})= 9.061-0.2x-0.237x^2-0.305x^3 \cr
-0.0283x^4 -y(0.0047-0.0221x -0.102x^2\cr
-0.0817x^3-0.00717x^4),
\end{eqnarray}
where $x \equiv$ log(\rr23) and $y \equiv$ log(\oo32).
These calibrations are represented in Fig.~\ref{calib} for
four typical values of the ionization parameter \oo32.

Figure~\ref{method} illustrates the procedure we adopt. A galaxy
lies on the {\em lower} branch (squares) if log(\nii/\halpha) $<
-1$ and log(\nii/\oii) $< -1.05$, or on the {\em upper} branch
(triangles) if log(\nii/\halpha) $> -1$ and log(\nii/\oii) $>
-0.8$. For galaxies with inconclusive values (i.e. $-1.05 <$
log(\nii/\oii) $< -0.8$), we follow the \nii/\halpha\ line ratio
criterion. A significant fraction of the galaxies (star symbols)
fall in the turnaround region of the O/H vs. \rr23\ relationship,
where both the \nii/\halpha\ and \nii/\oii\ diagnostics are
inconclusive. For these objects, \nii/\halpha\ indicates high
oxygen abundance (upper branch), whereas \nii/\oii\ suggests low
metallicity (lower branch) and we take the average value. The {\it
final} oxygen abundance values are listed in Table~\ref{magtab}
and are plotted in Fig.~\ref{calib}.

\subsection[]{Nitrogen abundance}

Nitrogen-to-oxygen abundance ratios (N/O) may be determined in the
absence of a measurement of the temperature-sensitive \oiiic\
emission line using the algorithm proposed by Thurston et al.
(1996). Again, this only requires the bright \nii, \oii\ and \oiii\ 
emission lines. The relationship, which is based on the same
premise as that between the oxygen abundance and \rr23, is
calibrated using photoionization models.

First, an estimate of the temperature in the \ntwo\
emission region (\tnii) is given by the empirical
calibration between \tnii\ and \rr23\ (Thurston et al. 1996):
\begin{eqnarray}
t_{\rm [N\,{\sc ii}]}=6065+1600(\log R_{\rm 23})+1878(\log R_{\rm
  23})^2\cr +2803(\log R_{\rm 23})^3.
\end{eqnarray}
The \ntwo\ temperature determined from the \rr23\ relation
can thus be used together with the observed strengths of \nii\
and \oii\ to determine the ionic abundance ratio N$^+$/O$^+$.
Pagel et al. (1992) gave the following formula based on
a five-level atom calculation:
\begin{eqnarray}
\log \frac{{\rm N}^+}{{\rm O}^+}=\log \frac{\rm [N\,{\sc ii}]\lambda
6584}{\rm [O\,{\sc ii}]\lambda 3727}+0.307-0.02\log t_{\rm
[N\,{\sc ii}]}\cr -\frac{0.726}{t_{\rm [N\,{\sc ii}]}},
\end{eqnarray}
where the \ntwo\ temperature is expressed in units of $10^4$
K. Finally, we assume that N/O $\equiv$ N$^+$/O$^+$.
There has been some discussions regarding the accuracy of this
assumption (e.g., Vila-Costas \& Edmunds 1993) but Thurston et al. (1996)
found through detailed modelling that this equivalence only introduces
small uncertainties in deriving N/O. The values of N/O with
their related uncertainties for our sample of UV galaxies
are listed in Table~\ref{magtab}.

\begin{figure}
 \vspace{-2.3cm}
 \epsfig{figure=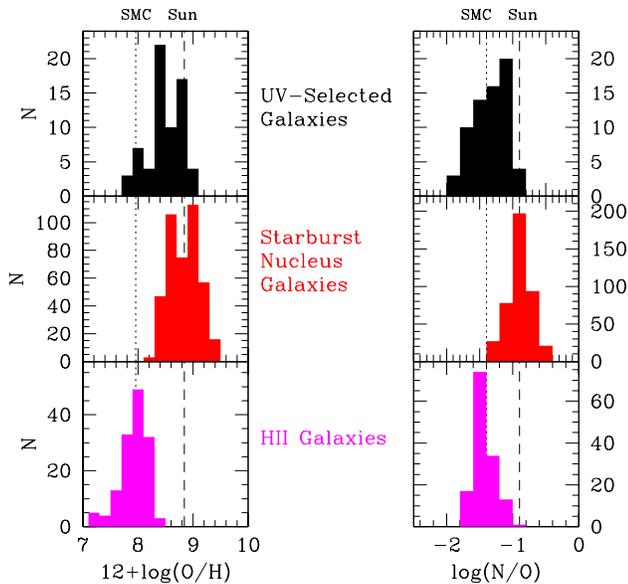,width=103mm}
 \caption{Distributions of {\it left}) the oxygen-to-hydrogen (\doh)
and {\it right}) nitrogen-to-oxygen (N/O) abundance ratios.
The distribution of the UV-selected galaxy sample is compared
to that of
Starburst Nucleus Galaxies and \hii\ galaxies (see
Fig.~\ref{novsoh} for references). Abundance values for the Sun
(dashed line) and for the SMC (dotted line) are indicated.}
 \label{abundhisto}
\end{figure}

\begin{table}
\caption{Mean values and standard deviation of O/H and N/O
abundance ratios for the UV-selected galaxies and comparison
samples of local star-forming galaxies.}
  \begin{center}
\begin{tabular}{lrrr}
\hline
Sample & N & 12+log(O/H) & log(N/O) \cr
\hline
UV galaxies & 68 & 8.49$\pm$ 0.04 & $-$1.35$\pm$ 0.03 \cr
SBNGs & 417 & 8.83$\pm$ 0.01 & $-$0.90$\pm$ 0.01 \cr
\hii\ galaxies & 139 & 7.91$\pm$ 0.02 & $-$1.43$\pm$ 0.01 \cr
\hline
\end{tabular}
  \end{center}
\label{abundstat}
\end{table}

\subsection[]{Systematic errors on N/O abundance ratios?}

The quoted errors on O/H and N/O abundance ratios listed in
Table~\ref{magtab} are uncertainties arising from measurement
errors only and do not take into account systematic calibration
errors. For the majority of the galaxies, the observational
uncertainties are indeed dominant ($\ga 0.2$ dex). Nonetheless,
it is appropriate to ask whether there may be systematic problems
with the abundance determinations.  Most notably, we may
underestimate the underlying Balmer absorption strength in some
galaxies. In this case, the extinction coefficient \cbeta\ would
be overestimated and the line strengths of ``blue'' emission lines
over-corrected. This effect might lead to an underestimate of the
N/O abundance ratio.

Balmer absorption strengths were directly measured in 55 galaxy
spectra ($\sim$ 80\% of the final sample), and are thus quite
reliable, but the assumed absorption equivalent width of 2 {\AA}\
for the other 13 galaxies could be incorrect. Fortunately, only
one of these galaxies exhibits a particularly low N/O value. We
conclude that our discussions in the following sections regarding
UV-selected galaxies with low N/O values are not strongly affected
by systematic errors on Balmer absorption.

\section[]{Analysis}
\label{analysis}

In this section, we use abundance ratios to constrain the physical
characteristics of the UV-selected galaxies. We compare their
chemical properties to those of various samples of normal and
star-forming galaxies in the local universe and at intermediate
redshifts.

As before, we consider local star-forming \hii\ galaxies and 
SBNGs (see Coziol et al. 1999 for the dichotomy). 
The \hii\ galaxy sample is a compilation
of irregular and blue compact dwarf galaxy samples from Kobulnicky
\& Skillman (1996) and Izotov \& Thuan (1999). The SBNG sample is
a merger of an optically-selected (Contini et al. 1998;
Consid{\`e}re et al. 2000) and far-infrared selected sample
(Veilleux et al. 1995).
Finally, we consider a small sample of intermediate-redshift
($0.1<z<0.5$) emission-line galaxies for which both O/H and
N/O abundance ratios are available (Kobulnicky \& Zaritsky 1999).

\subsection[]{The N/O versus O/H relationship}
\label{novsohrel}

\begin{figure*}
 \epsfig{figure=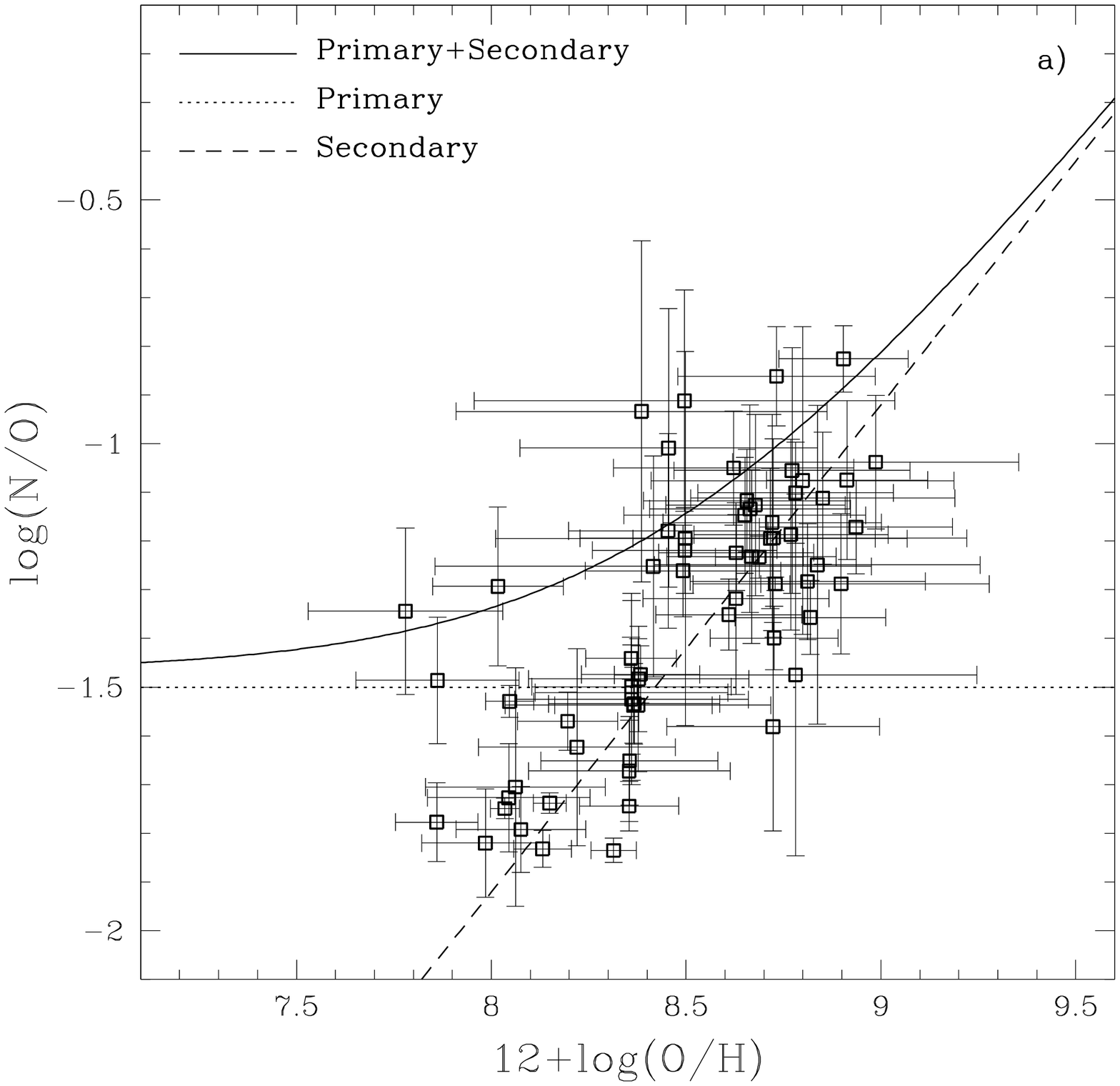,width=85mm}
 \epsfig{figure=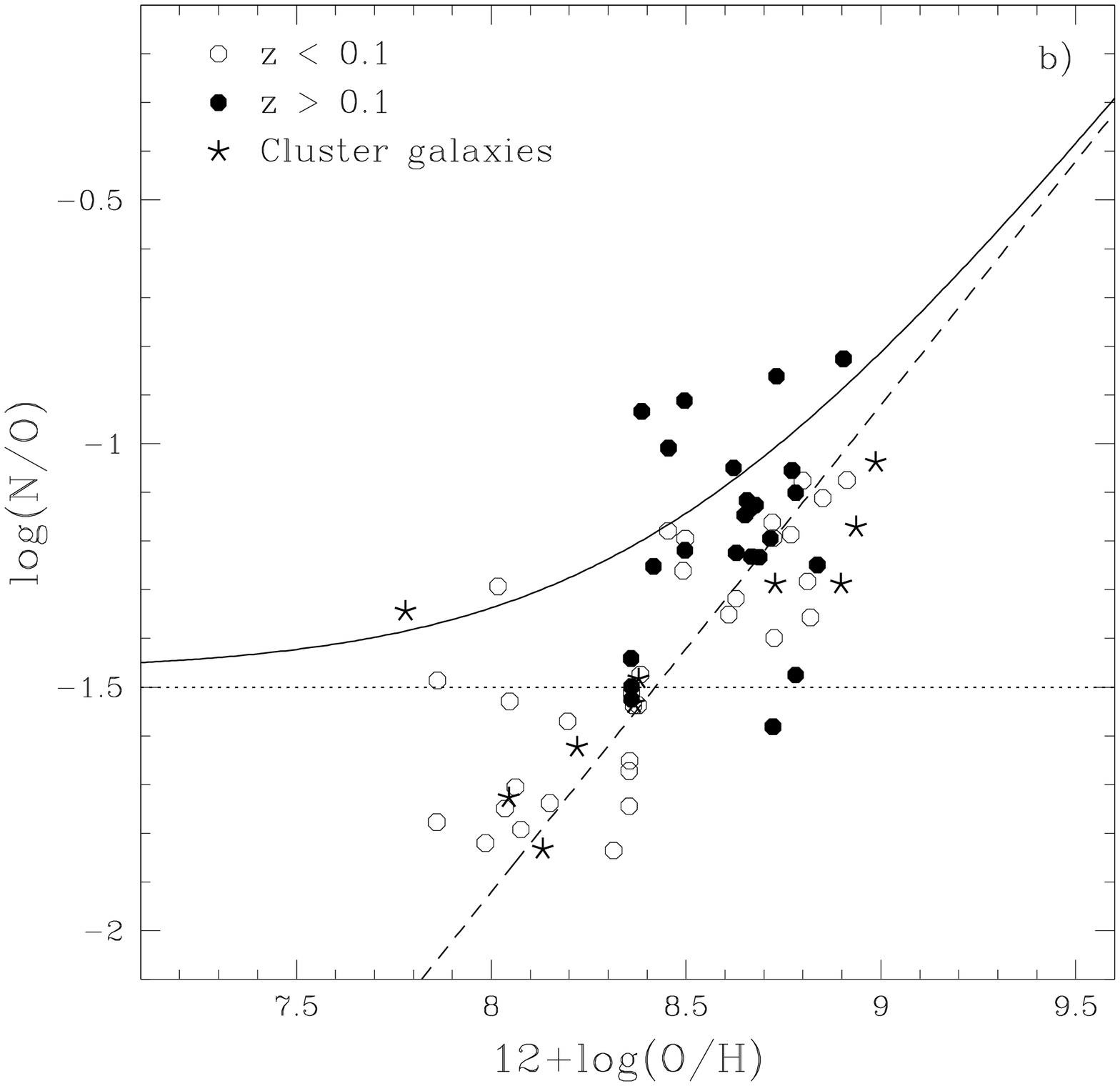,width=85mm}
 \caption{Nitrogen-to-oxygen (N/O) abundance ratio as a function
of oxygen abundance (\doh) for the UV-selected sample.
{\em a}) Abundance ratios are shown with error bars. Theoritical
curves for a {\em primary} (dotted line), a {\em secondary} (dashed
line), and a {\em primary + secondary} (solid line) production of
nitrogen (Vila-Costas \& Edmunds 1993) are shown.
{\em b}) A distinction is made between cluster (star symbols)
and field (circles) galaxies. Cluster galaxies are selected using the
redshift ranges $0.021 \leq z \leq 0.025$ for Coma, and
$0.019 \leq z \leq 0.023$ for Abell 1367. A further distinction
is made between field galaxies at low redshift ($z < 0.1$,
empty circles) and at intermediate redshift ($ z \geq 0.1$,
filled circles).
}
 \label{novsoherr}
\end{figure*}

The distributions of O/H and N/O abundance ratios for the UV-selected
galaxies are shown in Figure~\ref{abundhisto}, together with
comparison samples of local star-forming galaxies. Corresponding mean
values of O/H and N/O with their standard deviations are listed in
Table~\ref{abundstat}.

Our UV galaxies span a wide range of oxygen abundances, from \doh\
$\sim$ 7.7 ($\sim$ 0.1 \zsun) to 9.0 ($\sim$ \zsun). In terms of
metallicity, they are thus intermediate between low-mass \hii\
galaxies and massive SBNGs. The situation is quite similar for the
N/O abundance ratios; however the mean value is closer to that of
\hii\ galaxies (only 0.08 dex higher) than to that of the SBNGs
($\sim$ 0.45 dex lower). A closer look at Figure~\ref{abundhisto}
shows that a significant fraction of the UV galaxies has rather
low N/O abundance ratios (log(N/O) $< -1.7$). This appears to be
one of the most distinctive properties of the UV-selected sample.

\begin{figure*}
 \epsfig{figure=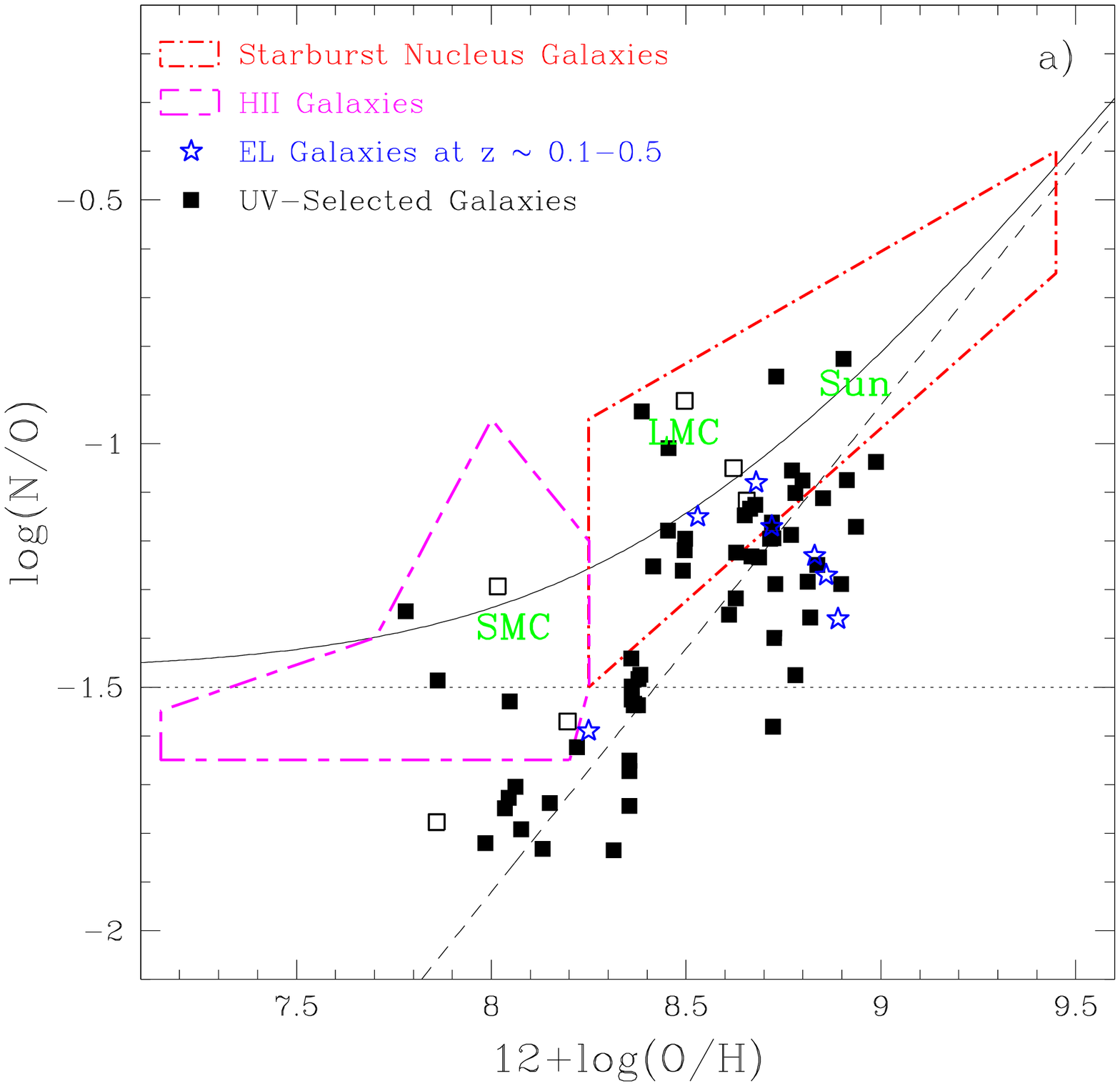,width=85mm}
 \epsfig{figure=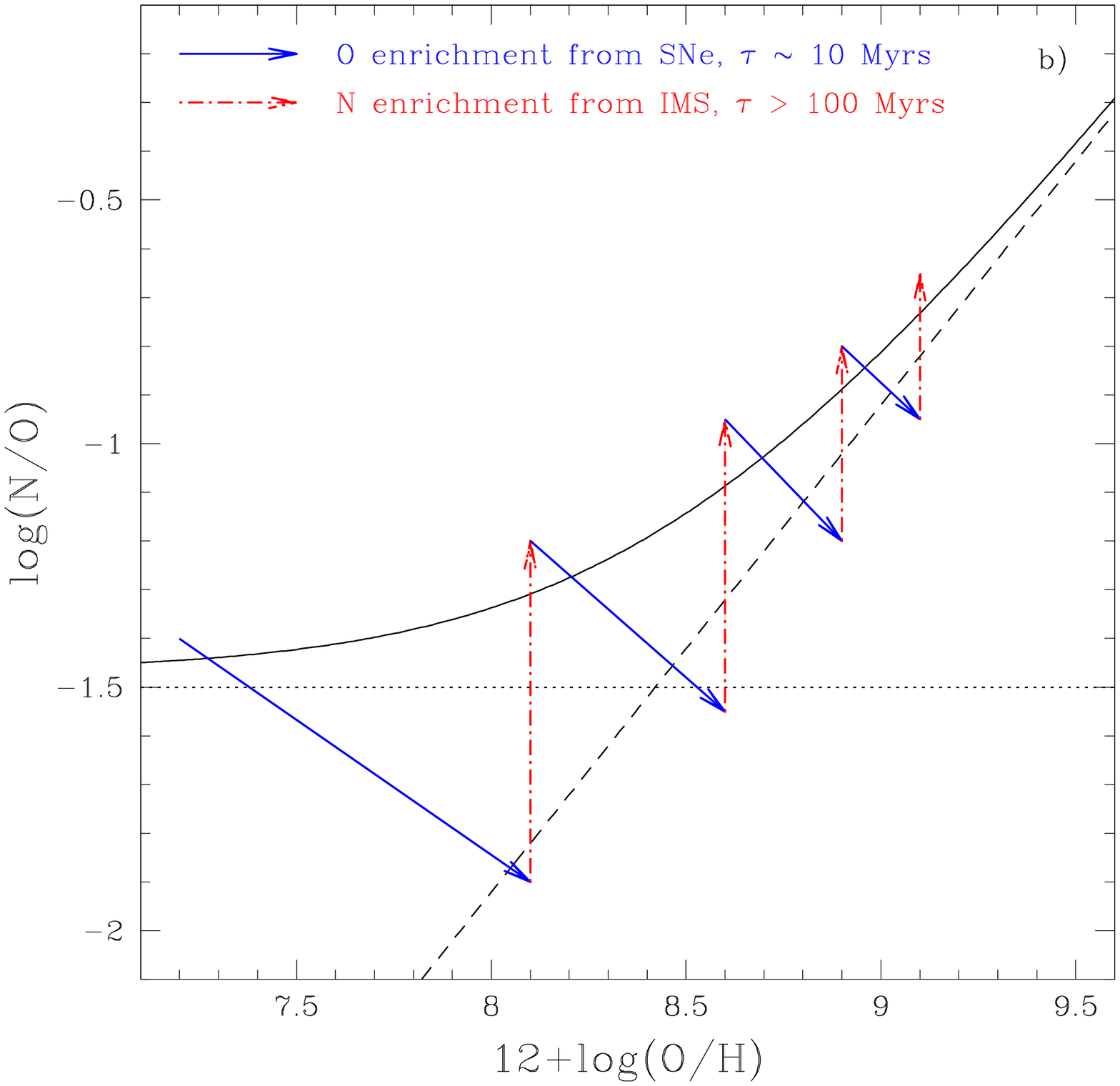,width=85mm}
 \caption{{\em a}) N/O versus \doh\ for the
UV-selected galaxies (squares) and various comparison
samples: Starburst Nucleus Galaxies (dot -- short dash line) selected
in the optical (Contini et al. 1998; Consid{\`e}re et al. 2000) or
in the far-infrared (Veilleux et al. 1995), \hii\ galaxies (short dash 
-- long dash line; Kobulnicky \& Skillman 1996; Izotov \& Thuan 1999),
and a small sample of emission-line galaxies at intermediate redshift
(stars; Kobulnicky \& Zaritsky 1999). Empty squares denote UV-selected
galaxies with two optical counterparts. Abundances ratios for the
Sun, the LMC and the SMC are also indicated.
{\em b}) Evolution model of N/O versus
\doh\ assuming a sequence of starbursts separated by quiescent periods
(see also Garnett 1990; Coziol et al. 1999). This
scenario assumes that each burst first produces oxygen
enrichment due to massive star evolution on short time-scales
($\tau \sim 10$ Myrs), followed by significant nitrogen enrichment
on longer time-scales ($\tau > 100$ Myrs) due to intermediate-mass star
evolution (see text for details).
}
 \label{novsoh}
\end{figure*}

In Figure~\ref{novsoherr} and \ref{novsoh}, we examine how the N/O
abundance ratio varies with O/H. The behaviour of N/O with
increasing metallicity offers clues about the chemical evolution
history of the galaxies and the stellar populations responsible
for producing oxygen and nitrogen.

The origin of nitrogen has been a subject of debate for some years.
The basic nucleosynthesis process is well understood.  Nitrogen is
thought to be synthesized in the CNO processing during hydrogen
burning. However the stars responsible remain uncertain.  If oxygen
and carbon are produced in previous generations, then nitrogen
produced in new stars should be proportional to the initial heavy
element abundance (i.e. {\it secondary} synthesis).  N/O should
increase linearly with O/H, and such a correlation is indeed observed
in high-metallicity (\doh\ $> 8.5$) \hii\ regions (see
Fig.~\ref{novsoh}). On the other hand, if oxygen and carbon are
produced in the same stars prior to the CNO cycle rather than in
previous generations, then nitrogen production is independent of the
initial heavy element abundance ({\it primary} synthesis).  This
regime seems to dominate at low metallicities (\doh\ $< 8.0$; e.g.,
Matteucci \& Tosi 1985; Matteucci 1986), for example in dwarf \hii\ 
galaxies where N/O exceeds the extrapolation of the linear trend
present at high abundances and does not correlate with O/H (see
Fig.~\ref{novsoh}).

Standard evolution models predict that secondary nitrogen
synthesis can occur in stars of all masses, while primary nitrogen
synthesis is usually thought to occur mainly in the convective
envelope of intermediate-mass stars ($\sim 3-8$ \msun) during the
AGB phase (e.g. Renzini \& Voli 1981; van den Hoek \& Groenewegen
1997; Henry, Edmunds \& K{\"o}ppen 2000).
However considerable progress has recently been made in including
rotational effects on the transport of elements and angular
momentum in stellar interiors (see review by Maeder \& Meynet
2000). In some massive stars the convective helium shell
penetrates into the hydrogen layer resulting in the production of
significant amounts of primary nitrogen (e.g., Maeder 2000). Thus
both intermediate-mass and massive stars are potential primary
nitrogen producers. Note that for intermediate-mass stars, the
most recent models (e.g. Marigo 2001) find a nitrogen
production similar to that of earlier models (e.g. Renzini \& Voli 1981)
at $Z \sim 0.2$ \zsun\ but which decreases with increasing metallicity.

A major problem with N/O abundance ratios has been to explain the
scatter in N/O at a given O/H (see Fig.~\ref{novsoh}). A natural
explanation might be a significant time delay between the release of
oxygen and that of nitrogen into the ISM (e.g., Edmunds \& Pagel 1978;
Garnett 1990; Pilyugin 1992, 1993, 1999; Vila-Costas \& Edmunds 1993;
Olofsson 1995b; Kobulnicky \& Skillman 1996, 1998; Coziol et al.
1999). Chemical evolution models of galaxies producing stars in short
bursts separated by long quiescent periods suggest that the dispersion
in N/O could be due to a delayed release of nitrogen produced in
low-mass longer-lived stars, compared to oxygen produced in massive,
short-lived stars.  The delayed-release hypothesis predicts that the
N/O ratio evolves significantly during a single cycle of star
formation followed by quiescence. Figure~\ref{novsoh} illustrates this
scenario. During a long period of quiescence, intermediate-mass star
evolution will significantly enrich the galaxy in nitrogen but not in
oxygen nor in any other Type II supernova product. At the end of a
long quiescent period, the N/O ratio should be high and level out
around \lno\ $\sim -1.5$ (e.g. Matteucci 1986; Olofsson 1995b), as a
result of intermediate-mass star evolution over the last few hundred
Myrs. During starburst however, N/O drops while O/H increases as the
most massive stars begin to die and supernovae release oxygen into the
ISM.
A few tens of Myrs after the burst, the massive O stars producing
oxygen will be gone. At this point, N/O should be minimal, with a
value roughly limited by the yields of O and secondary N from the
massive stars. Then N/O will rise again as intermediate-mass stars
begin to contribute to the primary and secondary production of
nitrogen, approaching a maximum value limited by the yields of N
and O from all stars. A galaxy undergoing successive starbursts
will thus oscillate between these two limits of N/O.

The vectors in Fig.~\ref{novsoh} illustrate this evolution for a
galaxy with an initial mass $10^8$ \msun, an initial metallicity
$Z \sim 0.02$ \zsun, and \lno\ $\sim -1.5$, which converts 1\% of its
mass into massive stars (see Garnett 1990). Massive star evolution
quickly moves the galaxy towards the lower right corner of the
diagram. Ejection of nitrogen from intermediate-mass stars then
increases N/O at constant O/H when the starburst has faded. A
second burst moves the galaxy back towards the lower right corner,
although to a lesser extent because of the higher abundance in the
ISM. In this picture, the N/O ratio can be used as a ``clock'' to
determine the time since the last major episode of star formation.
The delayed-release hypothesis predicts that galaxies                              
with high N/O ratios experienced a succession of starbursts                         
separated by rather long quiescent intervals, and are                               
preferentially picked out during a quiescent phase. On the other 
hand galaxies with low N/O ratios would have undergone shorter quiescent
phases and would be picked out at the end of a strong episode of 
star-formation, when high-mass stars  
release large amounts of O into the ISM, while N is still                           
withheld in lower-mass stars.
{\it Most of the UV-selected galaxies seem to 
fit this latter evolutionary scenario.}

\begin{figure*}
 \epsfig{figure=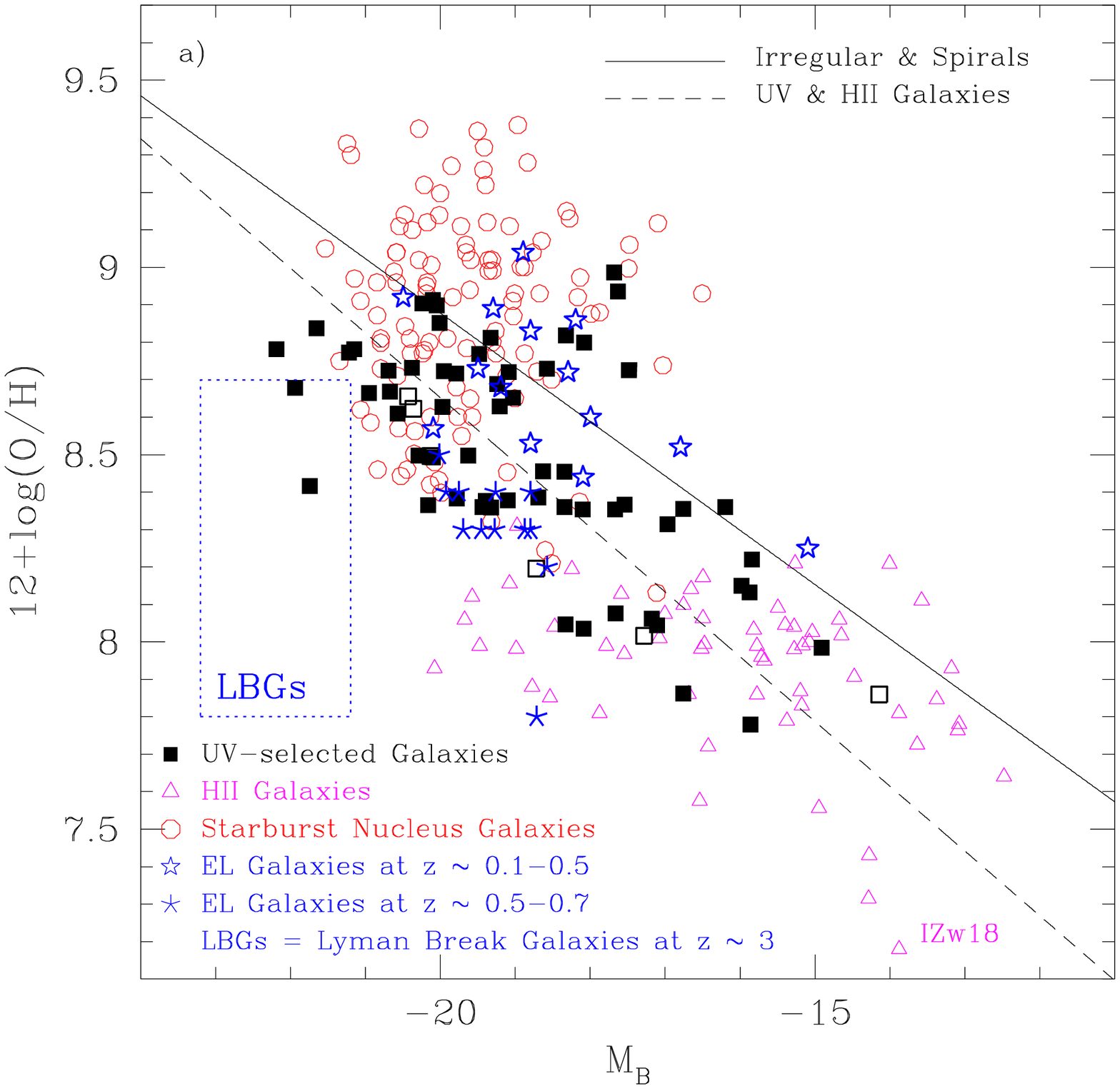,width=85mm}
 \epsfig{figure=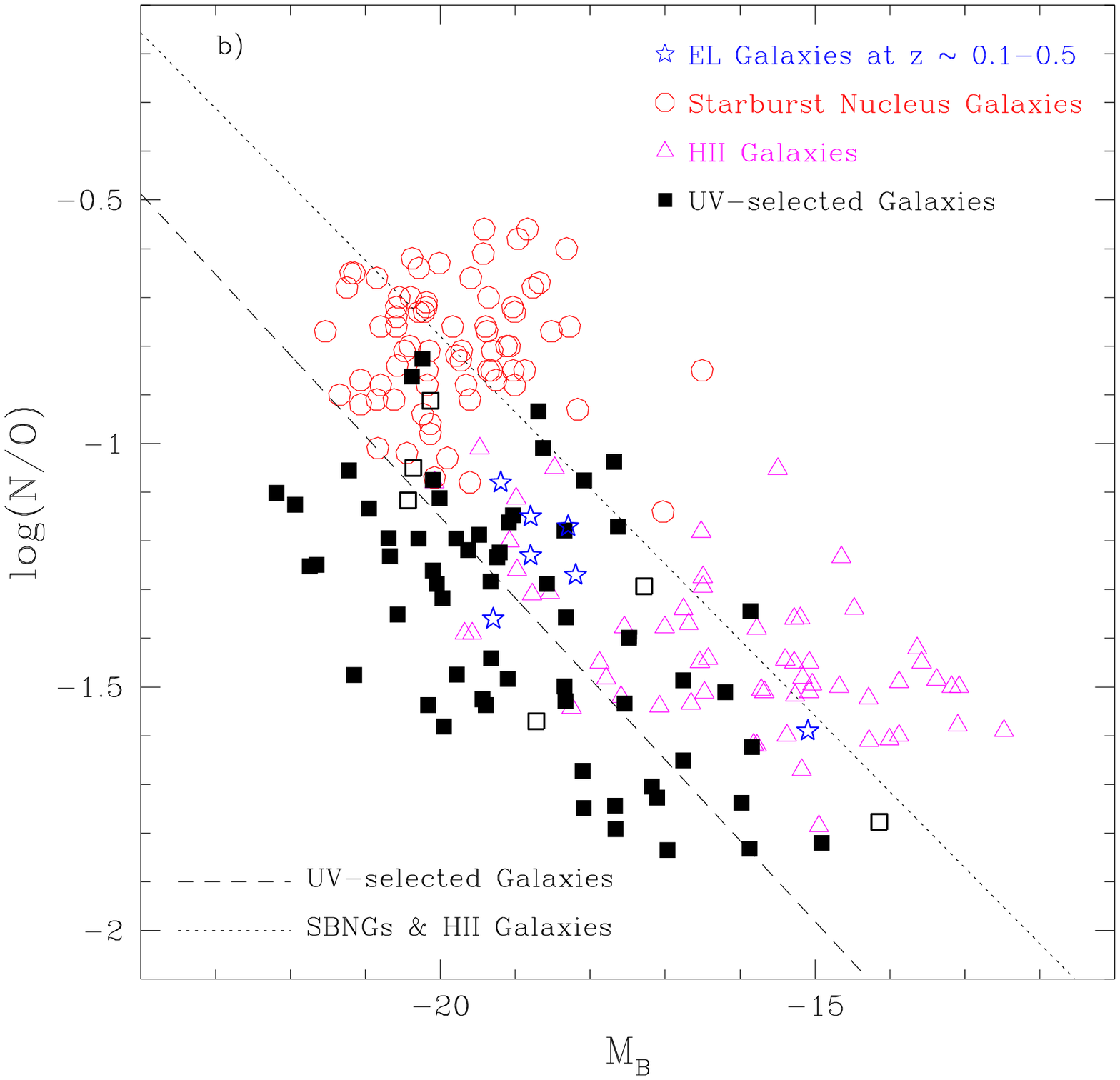,width=85mm}
 \caption{{\em a}) Metallicity--luminosity relation for the
UV-selected sample (squares) and for comparison samples of local, intermediate
and high redshift galaxies. The solid line is a linear
least-squares fit to local irregular and spiral galaxies (see Kobulnicky
\& Zaritsky 1999). The local star-forming galaxy samples are the same
as in Fig.~\ref{novsoh}: \hii\ galaxies (triangles) and SBNGs (circles).
Two samples of intermediate-redshift
galaxies (stars) are also shown for comparison: emission-line (EL)
galaxies at $z \sim 0.1-0.5$ (Kobulnicky \& Zaritsky 1999), and
luminous compact EL galaxies at $z \sim 0.5-0.7$ (Hammer et
al. 2001). The location of high-redshift ($z \sim 3$) Lyman break
galaxies is shown as a box encompassing the range of O/H and \mabs\
derived for these
objects (Pettini et al. 2001). The dashed line is a mean least-squares
fit to the UV-selected and \hii\ galaxies (correlation coefficient $r=-0.74$,
rms $=0.26$). The slope of this relation ($=-0.173\pm0.01$) is very
close to the one derived for local irregular and spirals galaxies (solid
line). UV-selected and \hii\ galaxies appear $2-3$ mag brighter than local
``normal'' galaxies of similar metallicity as might be expected if a strong
starburst has temporarily lowered their mass-to-light ratios.
{\em b}) N/O abundance ratio as a function of absolute $B$-band magnitude
for the UV-selected sample (squares) and the comparison samples
(see Fig.~\ref{novsoh} for references). The dashed line is a mean
least-squares fit to the UV-selected galaxies
(correlation coefficient $r=-0.56$, rms $=0.21$). The dotted line is
a mean least-squares fit to the SBNGs and \hii\ galaxies (correlation
coefficient $r=-0.79$, rms $=0.22$). These two relations have nearly
the same slope ($\sim -0.16$).
}
 \label{ohvsmabs}
\end{figure*}

We can also investigate whether the chemical properties of
UV-selected galaxies differ with environment -- field galaxies vs.
cluster galaxies -- and with redshift -- local galaxies ($z \la
0.1$) vs. intermediate-redshift galaxies ($z \ga 0.1$). 
In Figure~\ref{novsoherr}b we differentiate between cluster galaxies
(selected using the redshift ranges $0.021 \leq z \leq 0.025$ for
Coma, and $0.019 \leq z \leq 0.023$ for Abell 1367) and the
remaining sample, assumed to be field galaxies. 
No striking trends emerge, although there may be a
tendency for high-metallicity (\doh\ $> 8.7$) cluster galaxies to
have lower N/O abundance ratios than their field counterparts.
We also distinguish between low ($z < 0.1$) and intermediate 
redshift ($ z \geq 0.1$) field galaxies. 
The latter have relatively high oxygen
abundances (\doh\ $> 8.5$), some of them showing abundance
properties typical of SBNGs with rather high N/O abundance ratios
(log(N/O) $\ga -1$). This could be due to a succession of
starbursts over the last few Gyrs (e.g. Coziol et al. 1999;
Contini et al. 2000). The fact that UV galaxies at higher redshift
are more metal-rich objects is likely to be a selection effect arising 
from the well-known metallicity--luminosity relationship (see
Sect.~\ref{metlumrel}).

Finally, we discuss two major sources of uncertainties in the
delayed-release model. Firstly, the amount of gas and embedded
metals lost by a galaxy due to supernova-driven superwinds is
poorly constrained. If nitrogen is primary, then neither inflow of
unprocessed gas nor outflow of enriched gas will affect the N/O
ratio unless the outflow is different for nitrogen and oxygen
(e.g. Pilyugin 1993; Marconi, Matteucci \& Tosi 1994). If oxygen
is predominantly synthesized in massive stars while nitrogen is
produced in intermediate-mass stars, oxygen may be preferentially
ejected during supernova explosions. If nitrogen is secondary, the
N/O ratio is unaffected by any non-differential outflow, but can
be affected by unprocessed gas inflow (Serrano \& Peimbert 1983;
Edmunds 1990; K{\"o}ppen \& Edmunds 1999).

Secondly, the time-scale for heavy element enrichment is still not
well known. Since the nucleosynthetic products of massive stars
require longer than $\sim 10$ Myrs to mix with the surrounding ISM
(e.g., Tenorio-Tagle 1996; Kobulnicky \& Skillman 1997), there
will be some time lag ($> 10$ Myrs) between supernova explosions
and the appearance of fresh oxygen in the warm ionized gas. As
long as this time lag is no longer than the lifetime of
N-producing stars ($\sim$ some hundreds of Myrs), the N/O ratio
may serve as a useful ``clock'' measuring the time elapsed since
the last major burst of star formation. If however the time lag
required for massive star ejecta to cool down and mix with the ISM
is longer than a few $10^8$ yr, then the N/O ratio is unlikely to
reflect accurately the time since the most recent starburst accurately. One
test of the ``clock'' hypothesis would be to determine star
formation histories across a wide range of N/O abundance ratios.
Unfortunately, most of the galaxies in Figure~\ref{novsoh} are too
far away to directly investigate their star formation histories
from stellar colour-magnitude diagrams.

\subsection[]{Metallicity--luminosity relation}
\label{metlumrel}

We now study how the UV-selected galaxies compare with the
fundamental galaxy scaling relation between luminosity and
metallicity (e.g., Lequeux et al. 1979; Brodie \& Huchra 1991;
Skillman, Kennicutt, \& Hodge 1989; Zaritsky, Kennicutt \& Huchra
1994; Richer \& McCall 1995; Jablonka, Martin \& Arimoto 1996;
Coziol et al. 1997, 1998). This relation, which extends over $\sim
10$ magnitudes in luminosity and $\sim 2$ dex in metallicity,
presumably reflects the fundamental role that galaxy mass plays in
the chemical enrichment of the interstellar medium. The
metallicity-luminosity relationship is usually attributed to the
action of galactic superwinds; massive galaxies reach higher
metallicities because they have deeper gravitational potentials
better able to retain their gas against the building thermal
pressures from supernovae, whereas low-mass systems eject their
gas before high metallicities are attained (Larson 1974, De Young
\& Heckman 1994; Marlowe et al. 1995; MacLow \& Ferrara 1999;
Pilyugin \& Ferrini 1998, 2000; Silich \& Tenorio-Tagle 2001).

In Figure~\ref{ohvsmabs}, the UV-selected sample is compared to i)
local ``normal'' irregular and spiral galaxies, ii) nearby SBNGs
and \hii\ galaxies, and iii) samples of intermediate and high
redshift galaxies. The absolute $B$-band magnitudes of the
UV-selected galaxies span seven orders of magnitude, from \mabs\
$\sim -15$ to $-22$, with an average value \mabs\ $= -18.8$ 
(Fig.~\ref{histomag}a). In
Figure~\ref{ohvsmabs} we distinguish between UV galaxies with one
(filled squares) or two (empty squares) optical counterparts. The
solid line is a linear fit to the metallicity-luminosity relation
for local ``normal'' irregular and spiral galaxies (see Kobulnicky
\& Zaritsky 1999). Figure~\ref{ohvsmabs} includes samples of
nearby \hii\ galaxies (Kobulnicky \& Skillman 1996; Izotov \&
Thuan 1999) and SBNGs (Veilleux et al. 1995; Contini et al. 1998;
Consid{\`e}re et al. 2000) discussed earlier. Absolute magnitudes
for these were extracted from the LEDA database. Two samples of
intermediate-redshift galaxies are also shown for comparison:
emission-line galaxies at $z \sim 0.1-0.5$ (Kobulnicky \& Zaritsky
1999), and luminous compact emission-line galaxies at $z \sim
0.5-0.7$ (Hammer et al. 2001). The location of high-redshift ($z
\sim 3$) Lyman break galaxies is shown as a box encompassing the
range of O/H and \mabs\ derived for these objects (Pettini et al.
2001).

Figure~\ref{ohvsmabs}a clearly shows that both UV-selected and
\hii\ galaxies systematically deviate from the
metallicity-luminosity trend of local ``normal'' galaxies (solid
line). A mean least-squares fit (dashed line) to the UV-selected
and \hii\ galaxies (obtained by fitting both samples independently
and averaging the results) yields:
\begin{eqnarray}
12+\log({\rm O/H})=-0.173(\pm 0.01) M_{\rm B}+5.195(\pm 0.177),
\end{eqnarray}
with a correlation coefficient $r=-0.74$, and a rms deviation of
$0.26$. It is interesting to note that the slope of this relation
is similar to the one derived for local ``normal'' galaxies.
UV-selected and \hii\ galaxies thus appear $2-3$ mag brighter than
``normal'' galaxies of similar metallicity, as might be expected
if a strong starburst has temporarily lowered their mass-to-light
ratios. Note however that the departure of UV-selected galaxies
from the metallicity-luminosity trend of local ``normal'' galaxies
decreases with absolute magnitudes. Luminous (\mabs\ $\sim -20$)
UV-selected galaxies behave like low-metallicity SBNGs ($Z \la$ \zsun), 
which altogether do not follow the metallicity-luminosity relation. 
This could be
understood in the context of accretion of residual outlying gas
and/or small gas-rich galaxies. Following Struck-Marcell's (1981)
models, accretion of more gas than stars will result in a
steepening of the metallicity-luminosity relation, explaining the
behaviour of SBNGs and massive galaxies in general.

In Figure~\ref{ohvsmabs}a, we also show the locus of intermediate
($z < 1$) and high ($z \sim 3$) redshift objects. Whereas
emission-line galaxies with redshifts between $0.1$ and $0.5$
follow the metallicity-luminosity relation of ``normal'' galaxies,
there is a clear deviation for luminous and compact emission-line
galaxies at higher redshift ($z \sim 0.5-0.7$). The latter objects
better fit the metallicity-luminosity relation derived for
UV-selected and \hii\ galaxies. Hammer et al. (2001) argue that
these objects could be the progenitors of present-day spiral
bulges. The deviation is even stronger for LBGs at $z \sim 3$.
Even allowing for uncertainties in the determination of O/H and
\mabs, LBGs fall well below the metallicity-luminosity relation of
``normal'' local galaxies and have much lower abundances than
expected from this relation given their luminosities. The most
obvious interpretation (Pettini et al. 2001) is that LBGs have
mass-to-light ratios significantly lower than those of present-day
``normal'' galaxies. 

Figure~\ref{ohvsmabs}b shows the relation between N/O abundance
ratio and absolute blue magnitude for the UV-selected and the
comparison samples. Although the dispersion is higher than for the
metallicity-luminosity relation, there is a clear correlation
between \lno\ and \mabs. A mean least-squares fit (dashed line) to
the UV-selected galaxies yields:
\begin{eqnarray}
\log({\rm N/O})=-0.166(\pm 0.02) M_{\rm B}-4.472(\pm 0.275),
\end{eqnarray}
with a correlation coefficient $r=-0.56$, and a rms deviation of
$0.21$. A similar relationship appears if we merge together the
samples of SBNGs and \hii\ galaxies. A mean least-squares fit
(dotted line) to these nearby star-forming galaxies yields:
\begin{eqnarray}
\log({\rm N/O})=-0.156(\pm 0.01) M_{\rm B}-3.897(\pm 0.150),
\end{eqnarray}
with a correlation coefficient $r=-0.79$, and a rms deviation of
$0.22$. It is interesting that these relations have nearly the
same slope with a shift of $\sim 2-3$ mag in \mabs\ or $\sim 0.4$
dex in \lno. The rather low N/O abundance ratios in UV-selected
galaxies compared with other star-forming galaxies is obvious in
this plot. This supports the interpretation of the delayed-release
model explained in section~\ref{novsohrel}. We are witnessing
UV-selected galaxies at a special stage of their evolution, just
at the end of a powerful starburst which has suddenly enriched
the ISM in oxygen and temporarily lowered their mass-to-light
ratios. The fact that WR stars are not detected in the spectrum 
of UV-selected galaxies with low N/O values is another indication 
that the starburst is not young (age $\ga 10$ Myr) in these 
objects. On the other hand, the only galaxy with detected 
WR stars (see section~\ref{wr}), i.e. experiencing a rather 
young starburst, shows a high N/O abundance ratio (\lno\ $\sim -1.3$). 

\section{Conclusions}
\label{conclu}

We examined the chemical properties of a sample of UV-selected
intermediate-redshift ($0 \la z \la 0.4$) galaxies in the context of
their physical nature and star formation history. Our sample 
consists of 68 galaxies with heavy element abundance ratios, UV
and CCD $B$-band photometry. The main results are the following:

\begin{itemize}

\item {} Diagnostics based on emission-line ratios
show that {\em all} but one of the galaxies in our sample are powered
by hot, young stars rather than by an AGN. 

\item {} Massive WR stars, indicating the presence of a very young starburst, 
are detected in only one UV galaxy. In particular, no WR spectral features 
were identified in the spectrum of galaxies with extreme UV-optical colours, 
refuting the claim that these hot stars could be responsible for this 
UV excess (Brown et al. 2000).

\item {} UV-selected galaxies span a wide range of oxygen abundances, 
from $\sim$ 0.1 to 1 \zsun. They are intermediate between low-mass \hii\ 
galaxies and massive starburst nuclei, with no significant
distinction between cluster and field galaxies.

\item {} At a given oxygen abundance, the majority of the sample
shows rather low N/O abundance ratios compared to other local
samples of star-forming galaxies, such as \hii\ galaxies at low
metallicities and SBNGs at higher metallicities. This is one of
the most striking chemical properties of the UV-selected galaxies, 
allowing us to constrain their evolutionary stage in a new independent 
way (see below).

\item {} Like \hii\ galaxies, UV-selected galaxies systematically deviate
from the usual metallicity-luminosity relation. At a given metallicity
they appear to be 2-3 magnitudes brighter than ``normal''
quiescent galaxies.

\end{itemize}

We examined the above results in the context of the
``delayed-release'' chemical evolution model. According to this model,
most of the UV-selected galaxies recently experienced a powerful burst
of star formation which temporarily lowered both their mass-to-light
and N/O ratios.  At this stage, intermediate-mass stars did not have
enough time to evolve and release nitrogen, while the most massive
stars have died and released their newly synthesized oxygen into the
ISM through Type II supernova explosions. A significant number of
massive stars must nevertheless still be present to account for the
observed strong \halpha\ emission lines (S2000). 

If local UV galaxies represent scaled-down versions of the  
similarly UV-selected massive high-redshift LBGs, the present sample can be 
used to better understand the physical properties of these primordial 
galaxies. More quantitative
chemical evolution modeling of these UV galaxies' star formation 
history has been performed (Mouhcine \& Contini 2001).
Better quality spectra, near-infrared photometry as well as morphological 
informations should provide us with complementary observational constraints 
to probe deeper into the role of starbursts in galaxy formation and evolution.

\section*{Acknowledgments}

We thank R.\,Coziol, J.\,K{\"o}ppen, A.\,Lan{\c c}on, M.\,Mouhcine, 
and M.\,Pettini
for very fruitful discussions and useful suggestions. T.C. and
M.T. warmly acknowledge the hospitality of the Instituto National
de Astrofisica Optica y Electronica (INAOE), Mexico, during the
2000 Guillermo Haro International Program for Advanced Studies in
Astrophysics, during which this work was initiated.

\begin{table*}
\caption{The sample of UV-selected galaxies with measurements of
  chemical abundances. Column 1 gives an ID number, columns 2 and 3 
give the optical position
(RA and DEC, 1950) of the objects. Column 4 indicates the number
of optical counterparts (OC) on the POSS plates within 10\arcsec.
Columns 5 to 10 list emission-line ratios
used for standard diagnostic diagrams or as abundance indicators
(see text for details). Column 11 indicates if the object lies
on the {\em lower} (L) or on the {\em upper} (U) branch of the
O/H vs. \rr23\ relationship, or falls in the turnaround region (?).
Columns 12 and 13 give the oxygen abundance, \doh, derived from
the empirical calibrations (McGaugh 1991), on the {\em lower} and
on the {\em upper} branch respectively.}
{\scriptsize
\begin{tabular}{rrrrrrrrrrrrr}
\hline
 \# & RA (1950) & DEC (1950) & OC & [O\,{\sc iii}]/[O\,{\sc ii}] & [O\,{\sc iii}]/\hbeta & [N\,{\sc i
i}]/\halpha &
[S\,{\sc ii}]/\halpha & [N\,{\sc ii}]/[O\,{\sc ii}] & log(\rr23) & & O/H$_{\rm L}$ &
O/H$_{\rm U}$ \cr
(1) & (2) & (3) & (4) & (5) & (6) & (7) & (8) & (9) & (10) & (11) &
(12) & (13) \cr
\hline
 1& 13:06:32.16& $+$29:48:38.2& 1& $-$0.85$\pm$ 0.19& $-$0.70$\pm$ 0.20& $-$0.56$\pm$ 0.10& $-$0.83$\pm$ 0.10& $-$0.25$\pm$ 0.10&  0.25$\pm$ 0.15& U&  7.49&  8.99\cr
 2& 13:06:46.29& $+$29:43:37.9& 1& $-$0.42$\pm$ 0.25&  0.08$\pm$ 0.25& $-$1.00$\pm$ 0.07& $-$0.30$\pm$ 0.08& $-$1.04$\pm$ 0.07&  0.64$\pm$ 0.17& U&  7.94&  8.72\cr
 3& 13:06:50.23& $+$29:40:26.7& 2& $-$1.34$\pm$ 0.26& $-$0.83$\pm$ 0.34& $-$0.39$\pm$ 0.32&                & $-$0.45$\pm$ 0.30&  0.55$\pm$ 0.22& U&  8.05&  8.77\cr
 4& 13:06:30.66& $+$29:39:30.4& 1& $-$0.41$\pm$ 0.18&  0.32$\pm$ 0.25& $-$0.31$\pm$ 0.26&                & $-$0.59$\pm$ 0.25&  0.91$\pm$ 0.26& U&  8.35&  8.39\cr
 5& 13:06:07.74& $+$29:44:40.2& 1&  0.03$\pm$ 0.03&  0.52$\pm$ 0.04& $-$1.43$\pm$ 0.20& $-$0.97$\pm$ 0.05& $-$1.46$\pm$ 0.20&  0.87$\pm$ 0.03& L&  8.13&  8.51\cr
 6& 13:06:07.65& $+$29:36:35.6& 1& $-$1.03$\pm$ 0.24& $-$0.42$\pm$ 0.24& $-$0.45$\pm$ 0.21& $-$0.35$\pm$ 0.19& $-$0.61$\pm$ 0.22&  0.66$\pm$ 0.17& U&  8.15&  8.66\cr
 7& 13:07:05.26& $+$29:11:28.8& 2& $-$0.32$\pm$ 0.04&  0.31$\pm$ 0.06& $-$1.00$\pm$ 0.11& $-$0.49$\pm$ 0.05& $-$1.18$\pm$ 0.11&  0.84$\pm$ 0.05& L&  8.20&  8.50\cr
 8& 13:05:00.02& $+$29:40:04.3& 1&  0.06$\pm$ 0.03&  0.62$\pm$ 0.03& $-$1.42$\pm$ 0.11& $-$0.77$\pm$ 0.03& $-$1.53$\pm$ 0.11&  0.97$\pm$ 0.02& L&  8.31&  8.37\cr
 9& 13:03:49.57& $+$29:58:56.5& 1& $-$1.14$\pm$ 0.45& $-$0.64$\pm$ 0.46& $-$0.81$\pm$ 0.40&                & $-$0.87$\pm$ 0.37&  0.54$\pm$ 0.32& U&  8.03&  8.78\cr
10& 13:06:48.52& $+$29:02:35.2& 1& $-$0.18$\pm$ 0.06&  0.42$\pm$ 0.10& $-$1.12$\pm$ 0.37&                & $-$1.26$\pm$ 0.37&  0.89$\pm$ 0.09& L&  8.22&  8.46\cr
11& 13:04:13.54& $+$29:39:51.1& 2&  0.20$\pm$ 0.06&  0.54$\pm$ 0.12& $-$1.02$\pm$ 0.22&                & $-$0.90$\pm$ 0.22&  0.84$\pm$ 0.07& L&  8.02&  8.57\cr
12& 13:05:40.80& $+$29:14:43.1& 1& $-$0.64$\pm$ 0.74&  0.13$\pm$ 0.82& $-$0.91$\pm$ 0.82& $-$0.59$\pm$ 0.74& $-$1.23$\pm$ 0.80&  0.88$\pm$ 0.61& ?&  8.37&  8.40\cr
13& 13:04:37.54& $+$29:29:21.4& 1& $-$0.61$\pm$ 0.08& $-$0.33$\pm$ 0.09& $-$0.30$\pm$ 0.05& $-$0.32$\pm$ 0.04& $-$0.13$\pm$ 0.05&  0.41$\pm$ 0.06& U&  7.65&  8.90\cr
14& 13:03:37.14& $+$29:44:00.2& 1& $-$0.79$\pm$ 0.17& $-$0.05$\pm$ 0.20& $-$0.49$\pm$ 0.17& $-$0.29$\pm$ 0.14& $-$0.78$\pm$ 0.17&  0.83$\pm$ 0.16& U&  8.33&  8.45\cr
15& 13:03:48.44& $+$29:40:05.1& 1& $-$0.13$\pm$ 0.03&  0.27$\pm$ 0.04& $-$0.81$\pm$ 0.05& $-$0.52$\pm$ 0.03& $-$0.75$\pm$ 0.05&  0.71$\pm$ 0.04& U&  7.89&  8.69\cr
16& 13:06:11.39& $+$29:00:41.8& 1& $-$0.44$\pm$ 0.16& $-$0.05$\pm$ 0.26& $-$0.49$\pm$ 0.17& $-$0.25$\pm$ 0.15& $-$0.48$\pm$ 0.17&  0.56$\pm$ 0.26& U&  7.80&  8.80\cr
17& 13:05:55.86& $+$29:02:53.4& 1& $-$0.19$\pm$ 0.05&  0.50$\pm$ 0.06& $-$1.20$\pm$ 0.11& $-$0.60$\pm$ 0.07& $-$1.43$\pm$ 0.11&  0.96$\pm$ 0.04& L&  8.35&  8.35\cr
18& 13:02:53.21& $+$29:51:18.0& 1& $-$0.31$\pm$ 0.07&  0.22$\pm$ 0.11& $-$1.21$\pm$ 0.06& $-$0.83$\pm$ 0.12& $-$1.28$\pm$ 0.06&  0.76$\pm$ 0.09& L&  8.05&  8.61\cr
19& 13:02:43.94& $+$29:52:15.3& 1& $-$0.47$\pm$ 0.07&  0.09$\pm$ 0.10& $-$0.64$\pm$ 0.11&                & $-$0.75$\pm$ 0.10&  0.73$\pm$ 0.06& U&  8.06&  8.63\cr
20& 13:05:30.34& $+$29:03:16.3& 2& $-$0.47$\pm$ 0.11&  0.07$\pm$ 0.13& $-$0.54$\pm$ 0.09&                & $-$0.63$\pm$ 0.09&  0.71$\pm$ 0.09& U&  8.02&  8.66\cr
21& 13:04:23.62& $+$29:15:48.9& 1& $-$0.47$\pm$ 0.09&  0.20$\pm$ 0.09& $-$0.65$\pm$ 0.08& $-$0.17$\pm$ 0.08& $-$0.87$\pm$ 0.09&  0.84$\pm$ 0.08& U&  8.23&  8.49\cr
22& 13:03:28.24& $+$29:25:25.9& 1& $-$0.70$\pm$ 0.14& $-$0.17$\pm$ 0.19& $-$0.58$\pm$ 0.13&                & $-$0.65$\pm$ 0.13&  0.63$\pm$ 0.12& U&  7.99&  8.72\cr
23& 13:02:00.66& $+$29:47:57.0& 1& $-$0.36$\pm$ 0.05&  0.16$\pm$ 0.06& $-$0.60$\pm$ 0.07&                & $-$0.67$\pm$ 0.07&  0.72$\pm$ 0.09& U&  8.00&  8.65\cr
24& 13:04:00.52& $+$29:14:47.2& 2& $-$1.01$\pm$ 0.51& $-$0.28$\pm$ 0.46& $-$0.20$\pm$ 0.38&  0.20$\pm$ 0.34& $-$0.48$\pm$ 0.44&  0.78$\pm$ 0.34& U&  8.33&  8.50\cr
25& 13:04:50.67& $+$28:58:14.8& 1&  0.18$\pm$ 0.24&  0.63$\pm$ 0.29& $-$0.62$\pm$ 0.20&                & $-$0.68$\pm$ 0.22&  0.93$\pm$ 0.19& U&  8.20&  8.45\cr
26& 13:04:48.41& $+$28:54:49.3& 1& $-$0.33$\pm$ 0.18&  0.38$\pm$ 0.20& $-$0.92$\pm$ 0.18& $-$0.46$\pm$ 0.18& $-$1.17$\pm$ 0.18&  0.91$\pm$ 0.14& ?&  8.32&  8.40\cr
27& 13:04:19.78& $+$29:00:26.9& 2& $-$0.17$\pm$ 0.13&  0.35$\pm$ 0.13& $-$0.52$\pm$ 0.06& $-$0.51$\pm$ 0.06& $-$0.59$\pm$ 0.07&  0.75$\pm$ 0.09& U&  8.03&  8.62\cr
28& 13:04:28.46& $+$28:57:50.8& 1& $-$1.03$\pm$ 0.49& $-$0.23$\pm$ 0.50& $-$0.51$\pm$ 0.41&                & $-$0.86$\pm$ 0.46&  0.84$\pm$ 0.35& U&  8.41&  8.42\cr
29& 13:01:41.32& $+$29:39:36.8& 1& $-$0.51$\pm$ 0.08& $-$0.03$\pm$ 0.08& $-$0.30$\pm$ 0.09& $-$0.43$\pm$ 0.07& $-$0.32$\pm$ 0.09&  0.63$\pm$ 0.08& U&  7.91&  8.73\cr
30& 13:02:56.12& $+$29:18:53.6& 2&  0.10$\pm$ 0.04&  0.40$\pm$ 0.07& $-$1.47$\pm$ 0.35& $-$0.75$\pm$ 0.07& $-$1.31$\pm$ 0.36&  0.73$\pm$ 0.05& L&  7.86&  8.68\cr
31& 13:03:21.92& $+$29:08:18.0& 1& $-$0.40$\pm$ 0.05&  0.17$\pm$ 0.11& $-$1.20$\pm$ 0.18& $-$0.54$\pm$ 0.05& $-$1.34$\pm$ 0.18&  0.76$\pm$ 0.07& L&  8.08&  8.60\cr
32& 13:02:11.57& $+$29:25:28.0& 1& $-$0.34$\pm$ 0.21&  0.18$\pm$ 0.28& $-$1.21$\pm$ 0.23&                & $-$1.28$\pm$ 0.22&  0.79$\pm$ 0.19& L&  8.06&  8.59\cr
33& 13:02:14.19& $+$29:21:01.9& 1& $-$0.32$\pm$ 0.04&  0.40$\pm$ 0.04& $-$0.85$\pm$ 0.08& $-$0.65$\pm$ 0.05& $-$1.12$\pm$ 0.08&  0.94$\pm$ 0.04& ?&  8.36&  8.36\cr
34& 13:02:47.46& $+$29:07:09.6& 1& $-$0.40$\pm$ 0.12&  0.37$\pm$ 0.12& $-$0.88$\pm$ 0.14&                & $-$1.20$\pm$ 0.14&  0.94$\pm$ 0.09& ?&  8.36&  8.36\cr
35& 13:02:15.16& $+$29:14:25.5& 1& $-$0.23$\pm$ 0.09&  0.53$\pm$ 0.12& $-$1.05$\pm$ 0.15& $-$0.90$\pm$ 0.11& $-$1.36$\pm$ 0.15&  0.96$\pm$ 0.08& L&  8.35&  8.35\cr
36& 13:02:45.39& $+$29:04:56.9& 1& $-$1.15$\pm$ 0.41& $-$0.71$\pm$ 0.43& $-$0.61$\pm$ 0.27&                & $-$0.60$\pm$ 0.28&  0.48$\pm$ 0.30& U&  7.94&  8.84\cr
37& 13:00:49.01& $+$29:34:15.4& 1& $-$0.83$\pm$ 0.37& $-$0.11$\pm$ 0.38& $-$0.52$\pm$ 0.34&                & $-$0.78$\pm$ 0.35&  0.80$\pm$ 0.29& U&  8.28&  8.50\cr
38& 13:00:53.53& $+$29:31:42.1& 1& $-$1.09$\pm$ 0.22& $-$0.45$\pm$ 0.23& $-$0.63$\pm$ 0.18&                & $-$0.81$\pm$ 0.18&  0.68$\pm$ 0.16& U&  8.21&  8.63\cr
39& 13:03:48.16& $+$28:43:55.0& 1& $-$0.44$\pm$ 0.08&  0.28$\pm$ 0.09& $-$0.92$\pm$ 0.10& $-$0.58$\pm$ 0.07& $-$1.18$\pm$ 0.10&  0.89$\pm$ 0.06& ?&  8.32&  8.41\cr
40& 13:03:45.03& $+$28:43:35.2& 1& $-$0.67$\pm$ 0.05& $-$0.06$\pm$ 0.06& $-$0.72$\pm$ 0.05& $-$0.58$\pm$ 0.04& $-$0.88$\pm$ 0.05&  0.73$\pm$ 0.06& U&  8.12&  8.61\cr
41& 13:02:51.82& $+$28:53:38.2& 1& $-$0.46$\pm$ 0.08&  0.31$\pm$ 0.10& $-$0.88$\pm$ 0.09& $-$0.67$\pm$ 0.09& $-$1.20$\pm$ 0.10&  0.92$\pm$ 0.06& ?&  8.37&  8.37\cr
42& 13:00:59.29& $+$29:22:49.2& 1& $-$0.35$\pm$ 0.13&  0.03$\pm$ 0.20& $-$0.71$\pm$ 0.19&                & $-$0.63$\pm$ 0.17&  0.61$\pm$ 0.17& U&  7.80&  8.77\cr
43& 13:02:42.78& $+$28:54:32.0& 1& $-$0.25$\pm$ 0.08&  0.47$\pm$ 0.11& $-$1.07$\pm$ 0.13& $-$0.59$\pm$ 0.09& $-$1.34$\pm$ 0.12&  0.95$\pm$ 0.07& L&  8.35&  8.35\cr
44& 13:01:53.07& $+$29:07:07.0& 1&  0.08$\pm$ 0.01&  0.56$\pm$ 0.03& $-$1.36$\pm$ 0.11& $-$0.98$\pm$ 0.05& $-$1.38$\pm$ 0.11&  0.89$\pm$ 0.01& L&  8.15&  8.49\cr
45& 13:01:59.06& $+$29:04:42.8& 1& $-$0.70$\pm$ 0.06&  0.09$\pm$ 0.06& $-$0.77$\pm$ 0.07& $-$0.50$\pm$ 0.07& $-$1.12$\pm$ 0.07&  0.89$\pm$ 0.05& ?&  8.38&  8.38\cr
46& 13:02:05.41& $+$29:01:54.8& 1& $-$0.23$\pm$ 0.08&  0.24$\pm$ 0.11& $-$1.25$\pm$ 0.26&                & $-$1.36$\pm$ 0.26&  0.75$\pm$ 0.07& L&  7.99&  8.64\cr
47& 13:01:58.51& $+$29:01:55.0& 2&  0.24$\pm$ 0.09&  0.65$\pm$ 0.20& $-$0.91$\pm$ 0.55&                & $-$0.87$\pm$ 0.55&  0.94$\pm$ 0.12& U&  8.19&  8.46\cr
48& 13:02:25.99& $+$28:51:14.9& 1& $-$0.28$\pm$ 0.06&  0.34$\pm$ 0.11& $-$0.67$\pm$ 0.13&                & $-$0.83$\pm$ 0.13&  0.85$\pm$ 0.06& U&  8.19&  8.50\cr
49& 13:02:11.95& $+$28:53:43.2& 1& $-$0.64$\pm$ 0.10&  0.14$\pm$ 0.11& $-$0.80$\pm$ 0.10& $-$0.60$\pm$ 0.11& $-$1.13$\pm$ 0.11&  0.90$\pm$ 0.09& ?&  8.38&  8.38\cr
50& 13:02:41.80& $+$28:42:17.0& 1& $-$0.91$\pm$ 0.12& $-$0.42$\pm$ 0.12& $-$0.47$\pm$ 0.13& $-$0.73$\pm$ 0.15& $-$0.50$\pm$ 0.13&  0.55$\pm$ 0.09& U&  7.96&  8.78\cr
51& 13:02:33.29& $+$28:43:05.3& 1& $-$0.66$\pm$ 0.18& $-$0.15$\pm$ 0.18& $-$0.56$\pm$ 0.17&                & $-$0.62$\pm$ 0.15&  0.63$\pm$ 0.19& U&  7.97&  8.72\cr
52& 13:01:49.26& $+$28:48:37.7& 1& $-$0.03$\pm$ 0.10&  0.32$\pm$ 0.14& $-$1.00$\pm$ 0.10&                & $-$1.00$\pm$ 0.10&  0.70$\pm$ 0.10& L&  7.86&  8.70\cr
53& 13:01:07.18& $+$28:57:30.8& 1&  0.19$\pm$ 0.09&  0.40$\pm$ 0.23& $-$1.10$\pm$ 0.08&                & $-$0.86$\pm$ 0.08&  0.71$\pm$ 0.14& L&  7.78&  8.72\cr
54& 11:40:59.21& $+$20:46:04.6& 1& $-$1.03$\pm$ 0.18& $-$0.44$\pm$ 0.18& $-$0.46$\pm$ 0.09& $-$0.54$\pm$ 0.09& $-$0.60$\pm$ 0.11&  0.65$\pm$ 0.15& U&  8.11&  8.68\cr
55& 11:41:26.28& $+$20:39:30.9& 1& $-$0.61$\pm$ 0.14&  0.12$\pm$ 0.14& $-$0.88$\pm$ 0.16& $-$0.48$\pm$ 0.12& $-$1.15$\pm$ 0.17&  0.85$\pm$ 0.10& ?&  8.31&  8.45\cr
56& 11:41:13.29& $+$20:31:34.6& 1& $-$0.43$\pm$ 0.01&  0.04$\pm$ 0.01& $-$0.73$\pm$ 0.01& $-$1.02$\pm$ 0.02& $-$0.75$\pm$ 0.01&  0.64$\pm$ 0.01& U&  7.91&  8.73\cr
57& 11:43:04.16& $+$20:22:49.6& 1& $-$1.19$\pm$ 0.20& $-$0.58$\pm$ 0.21& $-$0.54$\pm$ 0.18&                & $-$0.70$\pm$ 0.17&  0.64$\pm$ 0.14& U&  8.19&  8.67\cr
58& 11:41:22.60& $+$20:27:44.9& 1& $-$1.09$\pm$ 0.20& $-$0.74$\pm$ 0.19& $-$0.68$\pm$ 0.14& $-$0.90$\pm$ 0.12& $-$0.59$\pm$ 0.15&  0.40$\pm$ 0.14& U&  7.82&  8.90\cr
59& 11:41:56.61& $+$20:23:02.9& 1& $-$1.07$\pm$ 0.16& $-$0.66$\pm$ 0.14& $-$0.49$\pm$ 0.12& $-$0.76$\pm$ 0.13& $-$0.46$\pm$ 0.13&  0.47$\pm$ 0.12& U&  7.86&  8.85\cr
60& 11:40:48.41& $+$20:25:49.3& 1& $-$0.26$\pm$ 0.02&  0.27$\pm$ 0.03& $-$1.02$\pm$ 0.08& $-$0.76$\pm$ 0.04& $-$1.09$\pm$ 0.08&  0.77$\pm$ 0.03& L&  8.05&  8.60\cr
61& 11:41:01.40& $+$20:21:04.4& 1&  0.00$\pm$ 0.01&  0.44$\pm$ 0.02& $-$1.35$\pm$ 0.07& $-$0.59$\pm$ 0.04& $-$1.34$\pm$ 0.07&  0.81$\pm$ 0.01& L&  8.03&  8.58\cr
62& 11:39:39.87& $+$20:19:34.2& 1& $-$0.55$\pm$ 0.08& $-$0.36$\pm$ 0.11& $-$0.71$\pm$ 0.06& $-$0.63$\pm$ 0.03& $-$0.45$\pm$ 0.06&  0.36$\pm$ 0.09& U&  7.53&  8.94\cr
63& 11:42:19.65& $+$20:09:28.8& 1& $-$1.02$\pm$ 0.46& $-$0.47$\pm$ 0.50& $-$0.53$\pm$ 0.35&                & $-$0.63$\pm$ 0.39&  0.60$\pm$ 0.34& U&  8.07&  8.72\cr
64& 11:40:20.70& $+$20:14:37.2& 1& $-$0.87$\pm$ 0.09& $-$0.45$\pm$ 0.09& $-$0.77$\pm$ 0.07& $-$0.71$\pm$ 0.07& $-$0.74$\pm$ 0.07&  0.52$\pm$ 0.07& U&  7.84&  8.82\cr
65& 11:40:03.61& $+$20:14:47.5& 1& $-$0.50$\pm$ 0.06& $-$0.02$\pm$ 0.07& $-$0.84$\pm$ 0.06& $-$0.69$\pm$ 0.06& $-$0.86$\pm$ 0.06&  0.64$\pm$ 0.05& U&  7.92&  8.73\cr
66& 11:40:18.41& $+$20:13:05.5& 1& $-$0.35$\pm$ 0.19&  0.42$\pm$ 0.22& $-$0.86$\pm$ 0.27&                & $-$1.17$\pm$ 0.27&  0.94$\pm$ 0.14& ?&  8.36&  8.36\cr
67& 11:42:27.23& $+$20:02:00.1& 1& $-$0.79$\pm$ 0.19& $-$0.52$\pm$ 0.19& $-$0.55$\pm$ 0.13& $-$0.64$\pm$ 0.12& $-$0.37$\pm$ 0.13&  0.40$\pm$ 0.16& U&  7.63&  8.91\cr
68& 11:41:26.36& $+$20:03:43.3& 1& $-$0.98$\pm$ 0.15& $-$0.52$\pm$ 0.14& $-$0.65$\pm$ 0.11& $-$0.48$\pm$ 0.10& $-$0.66$\pm$ 0.11&  0.52$\pm$ 0.10& U&  7.93&  8.81\cr
\hline
\end{tabular}

}
\label{abundtab}
\end{table*}

\begin{table*}
\caption{The sample of UV-selected galaxies with measurements of
  chemical abundances. Column 1 gives an ID number, columns 2 and 3 
give the optical position
(RA and DEC, 1950) of the objects. Column 4 indicates the number
of optical counterparts (OC) on the POSS plates within 10\arcsec.
Column 5 gives the redshift. Column 6 gives the extinction
coefficient \av. Columns 7 and 8
give the O/H and N/O abundances ratios with their uncertainties.}
{\scriptsize
\begin{tabular}{rrrrrrrr}
\hline
\# & RA (1950) & DEC (1950) & OC & z & $A_{\rm V}$ &
12+log(O/H) & log(N/O) \cr
(1) & (2) & (3) & (4) & (5) & (6) & (7) & (8) \cr
\hline
 1& 13:06:32.16& $+$29:48:38.2& 1& 0.023&  0.57$\pm$ 0.06&  8.99$\pm$ 0.37& $-$1.04$\pm$ 0.14\cr
 2& 13:06:46.29& $+$29:43:37.9& 1& 0.113&  0.85$\pm$ 0.07&  8.72$\pm$ 0.27& $-$1.58$\pm$ 0.21\cr
 3& 13:06:50.23& $+$29:40:26.7& 2& 0.243&  0.98$\pm$ 0.27&  8.77$\pm$ 0.30& $-$1.05$\pm$ 0.25\cr
 4& 13:06:30.66& $+$29:39:30.4& 1& 0.228&  0.52$\pm$ 0.15&  8.39$\pm$ 0.48& $-$0.93$\pm$ 0.35\cr
 5& 13:06:07.74& $+$29:44:40.2& 1& 0.024&  0.57$\pm$ 0.03&  8.13$\pm$ 0.07& $-$1.83$\pm$ 0.04\cr
 6& 13:06:07.65& $+$29:36:35.6& 1& 0.123&  1.85$\pm$ 0.19&  8.66$\pm$ 0.26& $-$1.13$\pm$ 0.21\cr
 7& 13:07:05.26& $+$29:11:28.8& 2& 0.063&  0.61$\pm$ 0.04&  8.20$\pm$ 0.13& $-$1.57$\pm$ 0.06\cr
 8& 13:05:00.02& $+$29:40:04.3& 1& 0.018&  0.78$\pm$ 0.03&  8.31$\pm$ 0.06& $-$1.84$\pm$ 0.03\cr
 9& 13:03:49.57& $+$29:58:56.5& 1& 0.335&  2.27$\pm$ 0.37&  8.78$\pm$ 0.47& $-$1.48$\pm$ 0.37\cr
10& 13:06:48.52& $+$29:02:35.2& 1& 0.021&  0.13$\pm$ 0.03&  8.22$\pm$ 0.25& $-$1.62$\pm$ 0.20\cr
11& 13:04:13.54& $+$29:39:51.1& 2& 0.090&  0.33$\pm$ 0.04&  8.02$\pm$ 0.17& $-$1.29$\pm$ 0.16\cr
12& 13:05:40.80& $+$29:14:43.1& 1& 0.080&  2.44$\pm$ 0.77&  8.38$\pm$ 1.02& $-$1.60$\pm$ 0.40\cr
13& 13:04:37.54& $+$29:29:21.4& 1& 0.167&  0.35$\pm$ 0.03&  8.90$\pm$ 0.17& $-$0.83$\pm$ 0.07\cr
14& 13:03:37.14& $+$29:44:00.2& 1& 0.051&  0.81$\pm$ 0.12&  8.45$\pm$ 0.26& $-$1.18$\pm$ 0.20\cr
15& 13:03:48.44& $+$29:40:05.1& 1& 0.182&  0.15$\pm$ 0.01&  8.69$\pm$ 0.11& $-$1.23$\pm$ 0.04\cr
16& 13:06:11.39& $+$29:00:41.8& 1& 0.056&  0.00$\pm$ 0.13&  8.80$\pm$ 0.39& $-$1.08$\pm$ 0.32\cr
17& 13:05:55.86& $+$29:02:53.4& 1& 0.039&  1.20$\pm$ 0.06&  8.35$\pm$ 0.13& $-$1.74$\pm$ 0.05\cr
18& 13:02:53.21& $+$29:51:18.0& 1& 0.024&  0.39$\pm$ 0.06&  8.05$\pm$ 0.21& $-$1.73$\pm$ 0.11\cr
19& 13:02:43.94& $+$29:52:15.3& 1& 0.185&  0.61$\pm$ 0.07&  8.63$\pm$ 0.20& $-$1.22$\pm$ 0.08\cr
20& 13:05:30.34& $+$29:03:16.3& 2& 0.248&  1.79$\pm$ 0.10&  8.66$\pm$ 0.26& $-$1.12$\pm$ 0.10\cr
21& 13:04:23.62& $+$29:15:48.9& 1& 0.062&  1.18$\pm$ 0.09&  8.49$\pm$ 0.25& $-$1.26$\pm$ 0.09\cr
22& 13:03:28.24& $+$29:25:25.9& 1& 0.242&  0.61$\pm$ 0.12&  8.72$\pm$ 0.35& $-$1.20$\pm$ 0.14\cr
23& 13:02:00.66& $+$29:47:57.0& 1& 0.223&  0.57$\pm$ 0.04&  8.65$\pm$ 0.31& $-$1.15$\pm$ 0.12\cr
24& 13:04:00.52& $+$29:14:47.2& 2& 0.285&  1.77$\pm$ 0.41&  8.50$\pm$ 0.54& $-$0.91$\pm$ 0.23\cr
25& 13:04:50.67& $+$28:58:14.8& 1& 0.247&  0.00$\pm$ 0.21&  8.45$\pm$ 0.38& $-$1.01$\pm$ 0.29\cr
26& 13:04:48.41& $+$28:54:49.3& 1& 0.040&  0.87$\pm$ 0.19&  8.36$\pm$ 0.25& $-$1.51$\pm$ 0.19\cr
27& 13:04:19.78& $+$29:00:26.9& 2& 0.113&  1.09$\pm$ 0.04&  8.62$\pm$ 0.31& $-$1.05$\pm$ 0.12\cr
28& 13:04:28.46& $+$28:57:50.8& 1& 0.388&  2.07$\pm$ 0.49&  8.42$\pm$ 0.56& $-$1.25$\pm$ 0.23\cr
29& 13:01:41.32& $+$29:39:36.8& 1& 0.167&  0.96$\pm$ 0.07&  8.73$\pm$ 0.25& $-$0.86$\pm$ 0.10\cr
30& 13:02:56.12& $+$29:18:53.6& 2& 0.018&  0.33$\pm$ 0.03&  7.86$\pm$ 0.11& $-$1.78$\pm$ 0.08\cr
31& 13:03:21.92& $+$29:08:18.0& 1& 0.026&  0.00$\pm$ 0.04&  8.08$\pm$ 0.17& $-$1.79$\pm$ 0.09\cr
32& 13:02:11.57& $+$29:25:28.0& 1& 0.090&  0.63$\pm$ 0.16&  8.06$\pm$ 0.23& $-$1.71$\pm$ 0.25\cr
33& 13:02:14.19& $+$29:21:01.9& 1& 0.180&  1.13$\pm$ 0.04&  8.36$\pm$ 0.12& $-$1.44$\pm$ 0.04\cr
34& 13:02:47.46& $+$29:07:09.6& 1& 0.178&  1.48$\pm$ 0.12&  8.36$\pm$ 0.30& $-$1.52$\pm$ 0.11\cr
35& 13:02:15.16& $+$29:14:25.5& 1& 0.025&  1.07$\pm$ 0.10&  8.35$\pm$ 0.26& $-$1.67$\pm$ 0.10\cr
36& 13:02:45.39& $+$29:04:56.9& 1& 0.186&  1.48$\pm$ 0.25&  8.84$\pm$ 0.42& $-$1.25$\pm$ 0.33\cr
37& 13:00:49.01& $+$29:34:15.4& 1& 0.083&  1.74$\pm$ 0.36&  8.50$\pm$ 0.49& $-$1.20$\pm$ 0.38\cr
38& 13:00:53.53& $+$29:31:42.1& 1& 0.082&  1.40$\pm$ 0.21&  8.63$\pm$ 0.24& $-$1.32$\pm$ 0.20\cr
39& 13:03:48.16& $+$28:43:55.0& 1& 0.068&  0.98$\pm$ 0.07&  8.36$\pm$ 0.20& $-$1.54$\pm$ 0.08\cr
40& 13:03:45.03& $+$28:43:35.2& 1& 0.068&  1.22$\pm$ 0.04&  8.61$\pm$ 0.19& $-$1.35$\pm$ 0.07\cr
41& 13:02:51.82& $+$28:53:38.2& 1& 0.022&  1.22$\pm$ 0.09&  8.37$\pm$ 0.22& $-$1.53$\pm$ 0.08\cr
42& 13:00:59.29& $+$29:22:49.2& 1& 0.083&  0.50$\pm$ 0.13&  8.77$\pm$ 0.25& $-$1.19$\pm$ 0.20\cr
43& 13:02:42.78& $+$28:54:32.0& 1& 0.018&  1.22$\pm$ 0.09&  8.35$\pm$ 0.23& $-$1.65$\pm$ 0.09\cr
44& 13:01:53.07& $+$29:07:07.0& 1& 0.025&  0.39$\pm$ 0.01&  8.15$\pm$ 0.04& $-$1.74$\pm$ 0.02\cr
45& 13:01:59.06& $+$29:04:42.8& 1& 0.027&  1.26$\pm$ 0.06&  8.38$\pm$ 0.15& $-$1.47$\pm$ 0.06\cr
46& 13:02:05.41& $+$29:01:54.8& 1& 0.027&  0.00$\pm$ 0.07&  7.99$\pm$ 0.16& $-$1.82$\pm$ 0.11\cr
47& 13:01:58.51& $+$29:01:55.0& 2& 0.018&  0.22$\pm$ 0.06&  8.46$\pm$ 0.25& $-$1.19$\pm$ 0.34\cr
48& 13:02:25.99& $+$28:51:14.9& 1& 0.253&  0.39$\pm$ 0.06&  8.50$\pm$ 0.24& $-$1.22$\pm$ 0.09\cr
49& 13:02:11.95& $+$28:53:43.2& 1& 0.022&  1.44$\pm$ 0.12&  8.38$\pm$ 0.28& $-$1.48$\pm$ 0.11\cr
50& 13:02:41.80& $+$28:42:17.0& 1& 0.219&  1.87$\pm$ 0.13&  8.78$\pm$ 0.25& $-$1.10$\pm$ 0.10\cr
51& 13:02:33.29& $+$28:43:05.3& 1& 0.070&  1.07$\pm$ 0.15&  8.72$\pm$ 0.28& $-$1.16$\pm$ 0.22\cr
52& 13:01:49.26& $+$28:48:37.7& 1& 0.027&  0.00$\pm$ 0.10&  7.86$\pm$ 0.21& $-$1.49$\pm$ 0.13\cr
53& 13:01:07.18& $+$28:57:30.8& 1& 0.022&  0.26$\pm$ 0.07&  7.78$\pm$ 0.25& $-$1.34$\pm$ 0.17\cr
54& 11:40:59.21& $+$20:46:04.6& 1& 0.169&  1.87$\pm$ 0.10&  8.68$\pm$ 0.23& $-$1.13$\pm$ 0.19\cr
55& 11:41:26.28& $+$20:39:30.9& 1& 0.040&  1.50$\pm$ 0.13&  8.38$\pm$ 0.34& $-$1.54$\pm$ 0.14\cr
56& 11:41:13.29& $+$20:31:34.6& 1& 0.021&  1.55$\pm$ 0.01&  8.73$\pm$ 0.04& $-$1.29$\pm$ 0.01\cr
57& 11:43:04.16& $+$20:22:49.6& 1& 0.383&  1.81$\pm$ 0.13&  8.67$\pm$ 0.22& $-$1.23$\pm$ 0.18\cr
58& 11:41:22.60& $+$20:27:44.9& 1& 0.022&  1.90$\pm$ 0.15&  8.90$\pm$ 0.38& $-$1.29$\pm$ 0.14\cr
59& 11:41:56.61& $+$20:23:02.9& 1& 0.024&  2.81$\pm$ 0.12&  8.85$\pm$ 0.34& $-$1.11$\pm$ 0.14\cr
60& 11:40:48.41& $+$20:25:49.3& 1& 0.071&  0.31$\pm$ 0.01&  8.05$\pm$ 0.06& $-$1.53$\pm$ 0.03\cr
61& 11:41:01.40& $+$20:21:04.4& 1& 0.068&  0.44$\pm$ 0.01&  8.03$\pm$ 0.04& $-$1.75$\pm$ 0.02\cr
62& 11:39:39.87& $+$20:19:34.2& 1& 0.020&  0.02$\pm$ 0.00&  8.94$\pm$ 0.25& $-$1.17$\pm$ 0.10\cr
63& 11:42:19.65& $+$20:09:28.8& 1& 0.082&  2.14$\pm$ 0.44&  8.72$\pm$ 0.50& $-$1.19$\pm$ 0.20\cr
64& 11:40:20.70& $+$20:14:37.2& 1& 0.024&  0.78$\pm$ 0.06&  8.82$\pm$ 0.19& $-$1.36$\pm$ 0.08\cr
65& 11:40:03.61& $+$20:14:47.5& 1& 0.024&  1.13$\pm$ 0.06&  8.73$\pm$ 0.16& $-$1.40$\pm$ 0.06\cr
66& 11:40:18.41& $+$20:13:05.5& 1& 0.132&  0.96$\pm$ 0.18&  8.36$\pm$ 0.26& $-$1.50$\pm$ 0.19\cr
67& 11:42:27.23& $+$20:02:00.1& 1& 0.069&  1.31$\pm$ 0.12&  8.91$\pm$ 0.21& $-$1.08$\pm$ 0.16\cr
68& 11:41:26.36& $+$20:03:43.3& 1& 0.016&  2.09$\pm$ 0.12&  8.81$\pm$ 0.30& $-$1.28$\pm$ 0.12\cr
\hline
\end{tabular}

}
\label{magtab}
\end{table*}

\bsp

\label{lastpage}


\begin{thebibliography}{99}

\bibitem{} Alloin D., Collin-Souffrin S., Joly M., Vigroux, L., 1979,
A\&A, 78, 200
\bibitem{} Brodie J.~P., Huchra J.~P., 1991, ApJ, 379, 157
\bibitem{} Brown W.~R., Kenyon S.~J., Geller M.~J., Fabricant D.~G.,
2000, ApJ, 540, L83
\bibitem{} Calzetti D., 1997, AJ, 113, 162 
\bibitem{} Carollo C.~M., Lilly S.~J., 2001, ApJ, 548, L153 
\bibitem{} Cole S., Lacey C.~G., Baugh C.~M., Frenk C.~S., 2000, MNRAS, 319, 168 
\bibitem{} Consid{\`e}re S., Coziol R., Contini T., Davoust, E.,
2000, A\&A, 356, 89
\bibitem{} Conti P.~S., 1991, ApJ, 377, 115
\bibitem{} Conti P.~S., Leitherer C., Vacca W.~D., 1996, ApJ, 461, L87
\bibitem{} Contini T., Consid{\`e}re S., Davoust E., 1998, A\&AS, 130,
  285
\bibitem{} Contini T., Coziol R., Consid{\`e}re S., Davoust E., Reyes
  R.~E.~C., 2000, in ``Building Galaxies: From the Primordial Universe
to the Present", 34th Moriond Meeting, F. Hammer et al. (Editions
Frontieres), p. 229 (astro-ph/9812410)
\bibitem{} Contini T., Davoust E., Consid{\`e}re S., 1995, A\&A, 303, 440
\bibitem{} Coziol R., Contini T., Davoust E., Consid{\`e}re S., 1997,
  ApJ, 481, L67
\bibitem{} Coziol R., Contini T., Davoust E., Consid{\`e}re S., 1998,
ASP Conf.~Ser.~147: Abundance Profiles: Diagnostic Tools for Galaxy
History, p. 219
\bibitem{} Coziol R., Reyes R.~E.~C., Consid{\`e}re S., Davoust E.,
Contini T., 1999, A\&A, 345, 733
\bibitem{} De Young D.~S., Heckman T.~M., 1994, ApJ, 431, 598
\bibitem{} Dopita M.~A., Evans I.~N., 1986, ApJ, 307, 431
\bibitem{} Dopita M.~A., Kewley L.~J., Heisler C.~A., Sutherland R.~S., 2000,
ApJ, 542, 224
\bibitem{} Edmunds M.~G., 1990, MNRAS, 246, 678
\bibitem{} Edmunds M.~G., Pagel B.~E.~J., 1978, MNRAS, 185, 77
\bibitem{} Edmunds M.~G., Pagel B.~E.~J., 1984, MNRAS, 211, 507
\bibitem{} Franx M., Illingworth G.~D., Kelson D.~D., van Dokkum
  P.~G., Tran K., 1997, ApJ, 486, L75
\bibitem{} Garnett D.~R., 1990, ApJ, 363, 142
\bibitem{} Gonz{\'a}lez Delgado R.~M., Leitherer C., Heckman T.~M.,
1999, ApJS, 125, 489
\bibitem{} Guseva N.~G., Izotov Y.~I., Thuan T.~X., 2000, ApJ, 531,
  776
\bibitem{} Hammer F., Gruel N., Thuan T.~X., Flores H., Infante L., 
2001, ApJ, 550, 570 
\bibitem{} Heckman T.~M., Robert C., Leitherer C., Garnett D.~R., van
  der Rydt F., 1998, ApJ, 503, 646
\bibitem{} Henry R.~B.~C., Edmunds M.~G., K{\"o}ppen J., 2000, ApJ, 541, 660
\bibitem{} Izotov Y.~I., Thuan T.~X., 1998, ApJ, 500, 188
\bibitem{} Izotov Y.~I., Thuan T.~X., 1999, ApJ, 511, 639
\bibitem{} Izotov Y.~I., Thuan T.~X., Lipovetsky V.~A., 1994, ApJ, 435, 647
\bibitem{} Izotov Y.~I., Thuan T.~X., Lipovetsky V.~A., 1997, ApJS,
  108, 1
\bibitem{} Jablonka P., Martin P., Arimoto N., 1996, AJ, 112, 1415
\bibitem{} Kewley L.~J., Heisler C.~A., Dopita M.~A., Lumsden S., 2001, 
ApJS, 132, 37 
\bibitem{} Kauffmann G., Charlot S., 1998, MNRAS, 294, 705 
\bibitem{} Kauffmann G., Charlot S., Balogh M.~L., 2001, Apj in press (astro-ph/0103130)
\bibitem{} Kobulnicky H.~A., Kennicutt R.~C., Pizagno J.~L., 1999, ApJ, 514,
544
\bibitem{} Kobulnicky H.~A., Koo D.~C., 2000, ApJ, 545, 712
\bibitem{} Kobulnicky H.~A., Skillman E.~D., 1996, ApJ, 471, 211
\bibitem{} Kobulnicky H.~A., Skillman E.~D., 1997, ApJ, 489, 636
\bibitem{} Kobulnicky H.~A., Skillman E.~D., 1998, ApJ, 497, 601
\bibitem{} Kobulnicky H.~A., Zaritsky D., 1999, ApJ, 511, 118
\bibitem{} K{\"o}ppen J., Edmunds M.~G., 1999, MNRAS, 306, 317
\bibitem{} Kunth D., Mas-Hesse J.~M., Terlevich E., Terlevich R.,
  Lequeux J., Fall S.~M., 1998, A\&A, 334, 11
\bibitem{} Larson R.~B., 1974, MNRAS, 169, 229
\bibitem{} Leitherer C., et al., 1999, ApJS, 123, 3
\bibitem{} Lequeux J., Rayo J.~F., Serrano A., Peimbert M.,
Torres-Peimbert S., 1979, A\&A, 80, 155
\bibitem{} Lilly S.~J., Le Fevre O., Crampton D., Hammer F., Tresse
  L., 1995, ApJ, 455, 50
\bibitem{} Lowenthal J.~D., et al., 1997, ApJ, 481, 673
\bibitem{} Mac Low M., Ferrara A., 1999, ApJ, 513, 142
\bibitem{} Madau P., Pozzetti L., Dickinson M., 1998, ApJ, 498, 106
\bibitem{} Maeder A., 2000, New Astronomy Review, 44, 291
\bibitem{} Maeder A., Meynet G., 1994, A\&A, 287, 803
\bibitem{} Maeder A., Meynet G., 2000, ARA\&A, 38, 143
\bibitem{} Marigo P., 2001, A\&A, 370, 194 
\bibitem{} Marconi G., Matteucci F., Tosi M., 1994, MNRAS, 270, 35
\bibitem{} Marlowe A.~T., Heckman T.~M., Wyse R.~F.~G., Schommer R.,
1995, ApJ, 438, 563
\bibitem{} Matteucci F., 1986, MNRAS, 221, 911
\bibitem{} Matteucci F., Tosi M., 1985, MNRAS, 217, 391
\bibitem{} McCall M.~L., Rybski P.~M., Shields G.~A. 1985, ApJS,
57, 1
\bibitem{} McGaugh S.~S., 1991, ApJ, 380, 140
\bibitem{} McGaugh S.~S., 1994, ApJ, 426, 135
\bibitem{} Meurer G.~R., Heckman T.~M., Calzetti D., 1999, ApJ, 521, 64
\bibitem{} Meurer G.~R., Heckman T.~M., Lehnert M.~D., Leitherer C.,
Lowenthal J., 1997, AJ, 114, 54
\bibitem{} Meynet G., 1995, A\&A, 298, 767
\bibitem{} Milliard B., Donas J., Laget M., Armand C., Vuillemin A.,
1992, A\&A, 257, 24
\bibitem{} Mouhcine M., Contini T., 2001, A\&A, submitted
\bibitem{} Olofsson K., 1995a, A\&AS, 111, 57
\bibitem{} Olofsson K., 1995b, A\&A, 293, 652
\bibitem{} Osterbrock D.~E., 1989, Astrophysics of Gaseous Nebulae
and Active Galactic Nuclei, University Science Books:Mill Valley CA
\bibitem{} Pagel B.~E.~J., Edmunds M.~G., Blackwell D.~E., Chun M.~S.,
\& Smith G., 1979, MNRAS, 189, 95
\bibitem{} Pagel B.~E.~J., Edmunds M.~G., Smith G., 1980, MNRAS,
193, 219
\bibitem{} Pagel B.~E.~J., Simonson E.~A., Terlevich R.~J., Edmunds M.~G., 
1992, MNRAS, 255, 325 
\bibitem{} Papovich C., Dickinson M., Ferguson H.~C., 2001, ApJ in
  press (astro-ph/0105087)
\bibitem{} Pettini M., Kellogg M., Steidel C.~C., Dickinson M.,
Adelberger K.~L., Giavalisco M., 1998, ApJ, 508, 539
\bibitem{} Pettini M., Shapley A.~E., Steidel C.~C., Cuby J.-G.,
  Dickinson M., Moorwood A.~F.~M., Adelberger K.~L., Giavalisco M.,
  2001, ApJ, in press (astro-ph/0102456)
\bibitem{} Pilyugin L.~S., 1992, A\&A, 260, 58
\bibitem{} Pilyugin L.~S., 1993, A\&A, 277, 42
\bibitem{} Pilyugin L.~S., 1999, A\&A, 346, 428
\bibitem{} Pilyugin L.~S., 2000, A\&A, 362, 325
\bibitem{} Pilyugin L.~S., 2001, A\&A, 369, 594
\bibitem{} Pilyugin L.~S., Ferrini F., 1998, A\&A, 336, 103
\bibitem{} Pilyugin L.~S., Ferrini F., 2000, A\&A, 354, 874
\bibitem{} Popescu C.~C., Hopp U., 2000, A\&AS, 142, 247
\bibitem{} Renzini A., Voli M., 1981, A\&A, 94, 175
\bibitem{} Richer M.~G., McCall M.~L., 1995, ApJ, 445, 642
\bibitem{} Schaerer D., Contini T., Kunth D., 1999, A\&A, 341, 399
\bibitem{} Schaerer D., Contini T., Pindao M., 1999, A\&AS, 136, 35
\bibitem{} Schaerer D., Guseva N.~G., Izotov Y.~I., Thuan T.~X., 2000,
  A\&A, 362, 53
\bibitem{} Schaerer D., Vacca W.~D., 1998, ApJ, 497, 618
\bibitem{} Seaton M.~J., 1979, MNRAS, 187, 73P
\bibitem{} Serrano A., Peimbert M., 1983, Revista Mexicana de Astronomia
y Astrofisica, 8, 117
\bibitem{} Silich S., Tenorio-Tagle G., 2001, ApJ, 552, 91
\bibitem{} Skillman E.~D., 1989, ApJ, 347, 883
\bibitem{} Skillman E.~D., Kennicutt R.~C., Hodge P., 1989, ApJ,
347, 875
\bibitem{} Somerville R.~S., Primack J.~R., Faber S.~M., 2001, MNRAS, 320, 
504 
\bibitem{} Steidel C.~C., Giavalisco M., Pettini M., Dickinson M.,
Adelberger K.~L., 1996, ApJ, 462, L17
\bibitem{} Steidel C.~C., Adelberger K.~L., Giavalisco M., Dickinson
  M., Pettini M., 1999, ApJ, 519, 1
\bibitem{} Struck-Marcell C., 1981, MNRAS, 197, 487
\bibitem{} Sullivan M., Treyer M.~A., Ellis R.~S., Bridges T.~J.,
Milliard B., Donas J., 2000, MNRAS, 312, 442
\bibitem{} Sullivan M., Mobasher B., Chan B., Cram L., Ellis R., Treyer M., 
Hopkins, A., 2001, ApJ, 558, 72 
\bibitem{} Telles E., Terlevich R., 1997, MNRAS, 286, 183
\bibitem{} Tenorio-Tagle G., 1996, AJ, 111, 1641
\bibitem{} Thurston T.~R., Edmunds M.~G., Henry R.~B.~C., 1996, MNRAS, 
283, 990 
\bibitem{} Treyer M.~A., Ellis R.~S., Milliard B., Donas J., Bridges
  T.~J., 1998, MNRAS, 300, 303
\bibitem{}  Vacca W.~D., Conti P.~S., 1992, ApJ, 401, 543
\bibitem{} van den Hoek L.~B., Groenewegen M.~A.~T., 1997, A\&AS, 123, 305
\bibitem{} van Zee L., Salzer J.~J., Haynes M.~P., O'Donoghue A.~A.,
Balonek T.~J., 1998, AJ, 116, 2805
\bibitem{} Veilleux S., Osterbrock D.~E., 1987, ApJS, 63, 295
\bibitem{} Veilleux S., Kim D., Sanders D.~B., Mazzarella J.~M.,
  Soifer B.~T., 1995, ApJS, 98, 171
\bibitem{} Vila Costas M.~B., Edmunds M.~G., 1993, MNRAS, 265, 199
\bibitem{} Williams R.~E. et al., 1996, AJ, 112, 1335
\bibitem{} Zaritsky D., Kennicutt R.~C., Huchra J.~P., 1994, ApJ,
420, 87
\end{thebibliography}
\end{document}